\documentclass[%
reprint,
 amsmath,amssymb,
aps,
pra,
]{revtex4-1}

\usepackage{graphicx}
\usepackage{dcolumn}
\usepackage{bm}
\usepackage[colorinlistoftodos]{todonotes}
\usepackage{soul}
\usepackage{hyperref}
\hypersetup{colorlinks=true, citecolor=blue, urlcolor=blue, linkcolor=blue}

\begin{document}


\title{Intrinsic timing jitter and latency in superconducting single photon nanowire detectors}

\author{J. P. Allmaras}
\email{jallmara@caltech.edu}
 \altaffiliation[Also ]{Department of Applied Physics, California Institute of Technology}
\affiliation{Jet Propulsion Laboratory, California Institute of Technology, Pasadena, California 91109, USA}
  
\author{A. G. Kozorezov}%
\affiliation{Department of Physics, Lancaster University, Lancaster, LA1 4YB, UK}

\author{B. A. Korzh}
\affiliation{Jet Propulsion Laboratory, California Institute of Technology, Pasadena, California 91109, USA}

\author{M. D. Shaw}
\affiliation{Jet Propulsion Laboratory, California Institute of Technology, Pasadena, California 91109, USA}

\author{K. K. Berggren}
\affiliation{Department of Electrical Engineering and Computer Science, Massachusetts Institute of Technology, Cambridge, Massachusetts 02138, USA}

\date{\today}

\begin{abstract}

We analyze the origin of the intrinsic timing jitter in superconducting nanowire single photon detectors (SNSPDs) in terms of fluctuations in the latency of the detector response, which is determined by the microscopic physics of the photon detection process. We demonstrate that fluctuations in the physical parameters which determine the latency give rise to the intrinsic timing jitter. We develop a general description of latency by introducing the explicit time dependence of the internal detection efficiency. By considering the dynamic Fano fluctuations together with static spatial inhomogeneities, we study the details of the connection between latency and timing jitter. We develop both a simple phenomenological model and a more general microscopic model of detector latency and timing jitter based on the solution of the generalized time-dependent Ginzburg-Landau equations for the 1D hotbelt geometry. While the analytical model is sufficient for qualitative interpretation of recent data, the general approach establishes the framework for a quantitative analysis of detector latency and the fundamental limits of intrinsic timing jitter. These theoretical advances can be used to interpret the results of recent experiments measuring the dependence of detection latency and timing jitter on photon energy to the few-picosecond level.
    
\end{abstract}

\pacs{Valid PACS appear here}
\maketitle


\section{\label{sec:Intro}Introduction}
\quad When an incident photon is absorbed by a current carrying superconducting nanowire, the superconductivity is locally suppressed in a nonequilibrium region known as a hotspot \cite{goltsman_picosecond_2001}.  The nonequilibrium dynamics of this hotspot are a topic of broad interest in superconducting detectors, but precise modeling of the physical process remains an open topic of research.  While there have been intense experimental \cite{engel_temperature-dependence_2013,renema_experimental_2014,lusche_effect_2014,renema_effect_2015,marsili_hotspot_2016,gaudio_experimental_2016,caloz_optically_2017} and theoretical \cite{bulaevskii_vortex-assisted_2012,zotova_intrinsic_2014,engel_detection_2015,engel_detection_2015-1,vodolazov_vortex-assisted_2015,kozorezov_quasiparticle_2015,vodolazov_single-photon_2017,kozorezov_fano_2017}  efforts to understand the details of the detection mechanism in SNSPDs, there is still debate over the most appropriate model for understanding this nonequilibrium process in different regimes of photon energy, bias current, and temperature. Considerable effort has been focused on understanding the internal efficiency of nanowire detectors as a means of validating detection models, but less attention has been given to the timing properties predicted by these models.

\quad An important property of SNSPD systems in practice is the timing uncertainty associated with each detection event, also known as the timing jitter.  There are numerous sources of timing jitter in SNSPD systems.  It is now understood that the principal contributions to the timing jitter come from electrical and amplifier noise \cite{zhao_intrinsic_2011, wu_improving_2017}, longitudinal geometric jitter due to the finite propagation speed of microwave signals along the length of the nanowire \cite{calandri_superconducting_2016,zhao_single-photon_2017}, timing jitter induced by local inhomogeneities in the nanowire \cite{oconnor_spatial_2011,cheng_inhomogeneity-induced_2017}, and intrinsic timing jitter originating from the microscopic physics of the detection process itself. In a theoretical study, the transverse geometric jitter was investigated by considering the variation in detection latency as a function of the transverse location of photon absorption across the nanowire \cite{wu_vortex-crossing-induced_2017}. An experimental study of the jitter associated with meandered \cite{sidorova_physical_2017} and straight \cite{sidorova_intrinsic_2018} nanowires found asymmetry in the jitter profile which was attributed to intrinsic effects, and more recently, an increase in timing jitter was measured in straight NbN nanowires with increasing magnetic field \cite{sidorova_timing_2018} which was qualitatively explained by the hotspot model \cite{vodolazov_single-photon_2017}. The same model is used to study the effects of transverse position dependence on timing jitter \cite{vodolazov_minimal_2018}. Experimental progress through the use of specialized short structures and low noise cryogenic amplifiers has enabled the measurement of record low timing jitter in niobium nitride (NbN) nanowires, as low as 2.7~ps FWHM \cite{korzh_demonstrating_2018}. The photon energy and temperature dependence of timing jitter observed in these measurements suggests that intrinsic effects which derive from the physics of the photon detection process are relevant.  The reduction in instrumental sources of timing jitter to the few-picosecond level has enabled a new type of experiment where the latency difference between photons of two energies is measured directly \cite{korzh_demonstrating_2018}.  By directly observing the latency distribution as a function of the physical parameters of the device, it is now possible to experimentally validate microscopic models of the photon detection process more directly. As timing jitter is a critical parameter in many applications, a detailed microscopic theory is an essential tool for engineering higher time resolution detectors in the future.

\quad We begin by introducing and discussing the general relationship between latency and timing jitter in SNSPDs in Section \ref{sec:TheoryII}, and develop a simple phenomenological theory.  In the presence of dynamic Fano fluctuations and static spatial non-uniformities, studying and understanding the latency distribution of the detector is sufficient to predict the intrinsic jitter which would be observed in an experiment. We make use of general analytical properties of the detector latency to demonstrate that the non-Gaussian instrument response function observed in recent experiments \cite{korzh_demonstrating_2018} can be attributed to the positive curvature of the latency vs energy relationship.  We analyze the simplified hotbelt detection model, its prediction capabilities, and its limitations.  In Section \ref{sec:TDGLModel}, we derive and discuss the detector latency using the generalized time-dependent Ginzburg-Landau (TDGL) equations, together with energy balance and current continuity equations, and we analyze the solutions of the one dimensional hotbelt model.  Section \ref{sec:Discussion} presents a general discussion of the results and directions for future study.

\section{\label{sec:TheoryII}Latency of SNSPD response and timing jitter}
\quad While the intrinsic latency of photon detection is an important parameter for understanding the performance of a device, it is challenging to measure directly because of the picosecond time scales involved. While the absolute latency of an SNSPD has still not been characterized directly, recent experiments \cite{korzh_demonstrating_2018} have measured the relative latency difference between detection events for photons of two different energies (1550 and 775~nm wavelengths) generated in the same optical path. These experiments were made possible by using a high switching current device and a low-noise cryogenic amplifier to reduce the impact of instrumental noise-induced jitter. Furthermore, the length of the active region of the device was reduced to 5~\(\mu\)m, which reduces the effect of timing jitter associated with the propagation delay of electromagnetic signals. Given an estimated transmission speed of \(\sim\)6~\(\mu\)m/ps \cite{zhao_single-photon_2017}, this geometric timing jitter is below 1~ps, and can be neglected.  Measurements of the relative detector latency add a valuable piece of information about the time scales of photon detection, which are important in understanding the detector performance in an application. Both dynamic and static fluctuations in the detector will affect the detector latency and manifest themselves in the shape of timing jitter distributions.  Therefore, understanding the latency of a detector and its fluctuations is key to predicting its timing jitter.

\begin{figure}
\includegraphics[width=\linewidth]{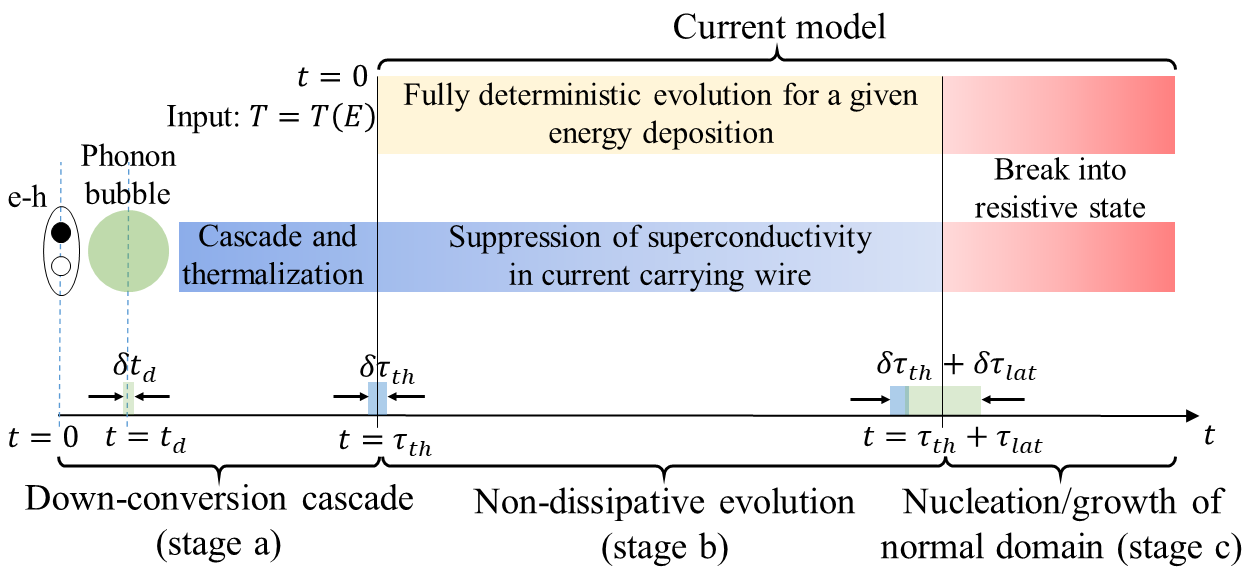} 
\caption{Schematic of SNSPD detection stages contributing to latency and intrinsic timing jitter.  We denote the down-conversion cascade as stage a), the non-dissipative suppression of superconductivity as stage b), and the nucleation and growth of the normal domain as stage c.)}
\label{F Latency Schematic}
\end{figure}

\quad The timescales of a typical photo-detection event in an SNSPD are shown schematically in Fig. \ref{F Latency Schematic}. Let's consider a photon of energy \(E_\lambda\) absorbed in a nanowire at an initial instance of time \(t = 0\)  resulting in the generation of a single electron-hole pair in a metal. The typical energy of the electron and hole of the pair is \(E_\lambda/2\). Due to the high energy of the initial excitations \((\geq 100~\text{meV})\) relative to the Fermi level they rapidly relax converting the deposited energy mostly into energetic phonon excitations \cite{kozorezov_quasiparticle-phonon_2000,kozorezov_fano_2017} described as the phonon bubble. This happens over the time scale \(\tau_d\) of a few tens of femtoseconds.  The variance \(\overline{(\delta{\tau_d}^2)}\) of this time is due to the distribution of initial energies of the primary e-h pair. It is of the order of the scattering time for and electron or hole emitting a single phonon with a frequency close to the Debye frequency, therefore \(|\delta\tau_d |\ll \tau_d\). Because \(\tau_d\) is so small, the duration of this time interval plays no role in any subsequent discussions.  It is convenient to consider the phonon bubble as a highly non-equilibrium initial distribution of elementary excitations (predominantly phonons, but with a small number of electronic excitations) with a total energy \(E_\lambda\) and radius \(2~\sqrt[]{D\tau_d}\), where \(D\) is a characteristic electron diffusion coefficient \cite{kozorezov_fano_2017, kozorezov_quasiparticle-phonon_2000}. This is the initial state, from which the evolution of non-equilibrium distribution of interacting phonons and electrons starts and proceeds as an energy down-conversion cascade with rapid multiplication of electron and phonon numbers \cite{vodolazov_single-photon_2017,kozorezov_fano_2017}. The down-conversion cascade ends at \(t = \tau_{th}\) with the electrons and phonons thermalized at a certain temperature.  The estimate of \(\tau_{th}\) in \cite{vodolazov_single-photon_2017} is in the range of 0.36 and 0.57 ps for WSi and NbN respectively for a 1.3 eV photon. This was derived under the assumption that diffusive expansion can be neglected for this down-conversion time. The estimates above serve as an indication of the order of magnitude of this process, and must be corrected to account for the material parameters, diffusion, and density of excitations in the excited volume (photon energy). Accounting for these changes results in a \(\tau_{th}\) on the order of a few ps for 1550 nm phonons (0.8 eV).  For the the typical case of a NbN SNSPD, this is smaller (much smaller) than the measured relative latencies, let alone the absolute latency, and \(\tau_{th} \ll \tau_{lat}\) as in Fig. \ref{F Latency Schematic}.  Similarly, the contribution to the timing jitter from the cascade stage a) is much smaller than the latency fluctuations over the non-dissipative stage b).  This outlines the main venue for the current paper, namely the study of the detector latency and its fluctuations over stages b) and c) as the dominant sources of the intrinsic timing jitter.  Summarizing, we discuss the model where a thermalized distribution of electrons and phonons (either in the form of a hotbelt or hotspot \cite{vodolazov_single-photon_2017}) at time \(\tau_{th}\), is considered as the initial state for the subsequent evolution of the superconducting system in a current carrying nanowire (top of Fig. 1). For each individual photon count, this state is characterized by an input energy \(E\) which denotes the amount of energy added to the superconducting system as a combination of electronic and phononic excitations in equilibrium at a temperature \(T\).  The subsequent evolution of the system is considered as fully deterministic with no other random factors affecting the onset of the resistive transition.  The period from \(\tau_{th}\) to \(\tau_{th} + \tau_{lat}\) in the notation of Fig. \ref{F Latency Schematic} we call the detector latency.  It is close to the absolute latency of the detector provided that \(\tau_{th} \ll \tau_{lat}\).  The timing jitter in this picture is associated with the fact that \(\tau_{lat}\) depends on the input energy \(E\) which fluctuates, depending on the actual energy loss from the individual cascade and hence from one click to another.

\quad We consider the single-photon detection regime and introduce the normalized-efficiency single-photon counting rate, \(PCR\), keeping its explicit dependence on time according to
\begin{equation}\label{PCR differential}
PCR(t,y,I_B,T_b,B,E)=\Theta\left[t- \tau_{lat}(y,I_B,T_b,B,E)\right]\text{,}
\end{equation}
where \(\tau_{lat}(y,I_B,T_b,B,E)\) is the SNSPD latency, depending on the transverse coordinate \(y\), bias current \(I_B\), bath temperature \(T_b\), external magnetic field \(B\) and energy absorbed by the superconductor electronic system and lattice \(E\). \(\Theta(t)\) is the Heaviside function. In this form, the \(PCR\) defines the probability of a detector click within the time interval [0, \(t\)] due to the absorption of energy \(E\). The energy \(E\) is less than the photon energy \(E_\lambda\) due to energy losses during the down-conversion cascade. These losses come from the escape of high energy phonons from the superconducting film to the substrate and potentially from coupling to plasmon modes in the superconductor. The details of this energy loss mechanism are not important in the following analysis. In this paper we will not explicitly discuss the magnetic field dependence, omitting the dependence of \(B\). This form also implicitly assumes perfect locality in the description of the photon absorption site and assumes deterministic evolution of the superconductor after photon absorption. Fluctuations are introduced by multiplying (\ref{PCR differential}) by the normalized Gaussian probability of energy deposition \(E\),
\begin{equation}\label{Gaussian}
P(E)=\frac{1}{\sqrt[]{2\pi}\sigma}e^{-\frac{\left(E-\bar{E}\right)^2}{2{\sigma}^2}}\text{,}
\end{equation}
with mean value \(\bar{E}\) and standard deviation \(\sigma\) to obtain
\begin{equation}\label{PCR general}
\begin{aligned}
PCR(t,y,&I_B,T_b,E_\lambda)={}\int_0^{E_\lambda}dE\, P(E)\Theta\left[I_B \right.\\
&\left.-I_{det}(y,T_b,E)\right]\Theta\left[t-\tau_{lat}( y,I_B,T_b,E)\right] \text{.}
\end{aligned}
\end{equation}
This is the general expression which is valid for both hotbelt (HB) and hotspot (HS) detection models discussed in the literature \cite{marsili_hotspot_2016,bulaevskii_vortex-assisted_2012,zotova_intrinsic_2014,kozorezov_quasiparticle_2015,kozorezov_fano_2017,vodolazov_single-photon_2017}.  We also introduce the photon counting rate averaged over the transverse coordinate of the absorption site $y$, 
\begin{equation}\label{PCR averaged over y}
\begin{aligned}
PCR&(t,I_B,T_b,E_\lambda)=\frac{1}{W}\int_{W/2}^{W/2}dy\int_0^{E_\lambda}dE\,P(E)\Theta\left[I_B-{} \right.\\ 
  &\left.I_{det}(y,T_b,E)\right]\Theta\left[t- \tau_{lat}( y,I_B,T_b,E)\right]
\end{aligned}
\end{equation}
for wire width \(W\), which is useful for describing the HS model. The first Heaviside function \(\Theta\left[I_B-I_{det}(y,T_b,E)\right]\) in the integrand ensures that an ideal detector  clicks with 100\% probability once the bias current exceeds the detection current \(I_{det}\) for a photon absorption at the site with transverse coordinate \(y\). The second Heaviside function \(\Theta\left[t-\tau_{lat}(y,I_B,T_b,E)\right]\)  allows the click to occur only after a deterministic interval of time, the detector response latency \(\tau_{lat}\), has elapsed. During this time following the absorption of a photon, a strongly non-equilibrium state of interacting quasiparticles and phonons evolves in time and space, suppressing the superconducting order parameter. At a certain stage, the superconducting state may become unstable. In the HS scenario \cite{engel_detection_2015,vodolazov_single-photon_2017}, either vortices enter from the edge of the wire or vortex-antivortex pairs are unbound inside the hotspot. In the HB scenario, phase slip lines are generated. If the bias current exceeds a certain threshold, which we call the detection current, energy dissipation in the current carrying nanowire results in the nucleation and growth of a normal domain through multiple vortex crossings or the formation of phase slip lines.

\subsection{\label{sec:RoleOfNU}Role of spatial non-uniformities}
\quad The standard deviation \(\sigma = \sqrt[]{{\sigma_{Fano}}^2+{\sigma_{n-u}}^2}\) describes the combined effect of Fano fluctuations (variance \({\sigma_{Fano}}^2\)) \cite{kozorezov_fano_2017} and spatial non-uniformity of the wire (variance  \({\sigma_{n-u}}^2\)). Spatial non-uniformity is assumed to originate from the spatial variation of parameters such as the local geometry (thickness of the wire) and local material parameters (critical temperature, density of states, and electron diffusivity). The use of the standard deviation \(\sigma\) in the form of the quadrature of the two statistically independent contributions as an approximation is justified, because the impact of the local non-uniformity can be described by fluctuations of the initial temperature in the excited volume for a fixed energy deposited in the nanowire. This in turn is formally equivalent to a homogeneous medium where local temperature fluctuations originate in fluctuating energy depositions with the appropriate variance. Below, we introduce a simple model to take into account the effect of spatial non-uniformity.

\quad The Fano fluctuations originate during the energy down-conversion cascade due to the energy loss from the thin nanowire film. While the Fano fluctuations are dynamic in nature, the spatial inhomogeneity of the wire is static in time. To derive the expression for \(\sigma_{n-u}\), we write down the energy conservation law \cite{vodolazov_single-photon_2017} in the form
\begin{equation}\label{conservation law}
\mathcal{E}_e(T_i) + \mathcal{E}_{ph}(T_i) = \mathcal{E}_e(T_b) + \mathcal{E}_{ph}(T_b) + E
\end{equation}
where \(\mathcal{E}_e(T_i)\) and \(\mathcal{E}_{ph}(T_i)\) are the energies for the equilibrated distributions of electrons and phonons at temperature \(T_i\) some short time following photon absorption and \(E\) is the energy gained by the system following the down-conversion process. For temperatures exceeding \(T_c\) after the cascade, the energy of the system can be expressed as \(\displaystyle{\mathcal{E}_e(T) + \mathcal{E}_{ph}(T) = {E}_cV_i\left[\frac{\pi^2}{12}\left(\frac{T}{T_c}\right)^2 + \frac{\pi^4}{15\gamma}\left(\frac{T}{T_c}\right)^4\right]}\), where \({E}_c=4N(0){k_B}^2{T_c}^2\) is the energy density of the system, \(N(0)\) is the single-spin electron density of states at the Fermi level in the normal state, \(V_i\) is the initial volume, \(\gamma=8\pi^2/5\left(C_e(T_c)/C_{ph}(T_c)\right)\) and \(C_e\) and \(C_{ph}\) are electron and phonon heat capacities respectively. For the HS model \cite{vodolazov_single-photon_2017}, \(V_i = 4{\xi_c}^2d\), where \({\xi_c}^2=\hbar£D/k_BT_c\), and \(d\) is the nanowire thickness. For the HB model, \(V_i = LWd\), where \(L\) is the length of a rectangular hotbelt along the nanowire of width \(W\). The only parameter in (\ref{conservation law}) which may exhibit spatial fluctuations is the energy \(E_cV_i\), through the combined variation of the electronic density of states, critical temperature, coherence length, and thickness of the wire, \(E_cV_i = \overline{E_cV_i} + \delta\left(E_cV_i\right)\), where the overline denotes the mean value and \(\delta\left(E_cV_i\right)\) is the energy fluctuation with zero mean value. Note that if the spatial fluctuations are characterized by a sufficiently small length scale \(r \ll L,W\), their effects may be strong in the HS case but will be substantially self-averaged and weakened for the HB scenario. In the presence of independent fluctuations of the energy deposition \(\delta E \) due to the Fano effect and the spatial non-uniformity \(\delta\left(E_cV_i\right)\), the solution of (\ref{conservation law}) yields fluctuations of the initial temperature in the excited volume with the standard deviation \cite{kozorezov_fano_2017}
\begin{equation}\label{T standard dev}
\begin{aligned}
\overline{\left(\delta T_i\right)^2} \propto {}& {\sigma_{Fano}}^2 + \bar{E}^2\overline{\delta\left(E_cV_i\right)^2}/ \overline{E_cV_i}^2 = \\
&{\sigma_{Fano}}^2 + \chi^2{E_\lambda}^2\overline{\delta\left(E_cV_i\right)^2}/ \overline{E_cV_i}^2 = \\
&{\sigma_{Fano}}^2 + {\sigma_{n-u}}^2 \\
{\sigma}^2 ={}& {\sigma_{Fano}}^2 + {\sigma_{n-u}}^2 = F_{\text{eff}}\varepsilon E_\lambda + \chi^2\Xi^2{E_\lambda}^2 = \\
&{\sigma_{Fano}}^2 \left(1 + a E_\lambda / E_{0} \right),
\end{aligned}
\end{equation}
where
\begin{equation}\label{n-u}
\begin{aligned}
\sigma_{n-u} = \chi\Xi{E_\lambda} \text{ and } \Xi^2 = \overline{\delta\left(E_cV_i\right)^2}/ \overline{E_cV_i}^2 
\end{aligned}
\end{equation}
and \(\chi\) is the mean fraction of the photon energy deposited in the nanowire. In the last line of expression (\ref{T standard dev}), the dimensionless constant \(a = \chi^2\Xi^2E_0/(F_{\text{eff}}\varepsilon)\) parametrizes the contribution of spatial non-uniformities relative to Fano fluctuations, and $E_{0}$ is the energy of a reference photon. As seen from the definitions of the Fano fluctuations \cite{kozorezov_fano_2017} and spatial non-uniformities, the respective variances scale differently with photon energy, \(\sigma_{Fano}\sim\sqrt[]{E_\lambda} \) and \(\sigma_{n-u}\sim{E_\lambda}\). Therefore, if experiments cover a broad range of wavelengths it cannot be excluded that Fano fluctuations dominate the jitter at the long wavelength end of the spectrum, while spatial non-uniformities become more important for shorter wavelengths.  This is the typical situation in single photon superconducting tunnel junction detectors, where the Fano fluctuation limit of spectral resolution is reached for energies below the hard ultraviolet, and spatial inhomogeneities become important at higher energies \cite{kozorezov_electron_2007}.

\subsection{\label{sec:HSModel}Timing jitter: hotspot model}
\quad The time-dependent photon counting rate \(PCR(t,y,I_B,T_b,E_\lambda)\) gives the probability of a detector click due to the absorption of a single photon within the time interval \(\left[0,t\right]\). Correspondingly, \(H(t,y,I_B,T_b,E_\lambda)dt = \frac{dPCR(t,y,I_B,T_b,E_\lambda)}{dt}dt\) is the probability of a single photon detection process within the time interval \(\left[t,t+dt\right]\), where \(H(t,y,I_B,T_b,E_\lambda) = \frac{dPCR(t,y,I_B,T_b,E_\lambda)}{dt}\) is the click probability density, also known as the instrument response function (IRF) for single photon absorption events.  To calculate the IRF, which is often the observed quantity in an experiment, we define the quantity \(E(t, y,I_B,T_b)\) which represents the energy transferred to the superconductor that corresponds to a latency \(t\) according to the single-valued solution of 
\begin{equation}\label{E t}
t - \tau_{lat}( y,I_B,T_b,E)= 0\text{.}
\end{equation}
Single-valuedness follows from the fact that the detector latency can only be a monotonically decreasing function of the energy deposition, \(\displaystyle{\frac{\partial \tau_{lat}(y,I_B,T_b,E)}{\partial E} < 0}\). Using the definition of \(E(t, y,I_B,T_b)\), we re-write expression (\ref{PCR general}) in the form
\begin{equation}\label{PCR t}
\begin{aligned}
PCR&(t,y,I_B,T_b,E_\lambda)=\int_0^{E_\lambda}dE\, P(E)\Theta\left[I_B-{}\right.\\ 
  &\left.I_{det}(y,T_b,E)\right]\Theta\left[E-E(t, y,I_B,T_b)\right] \\
  {}=&\int_{E(t, y,I_B,T_b)}^{E_\lambda}dE\, P(E)\Theta\left[I_B-I_{det}(y,T_b,E)\right] \text{.}
\end{aligned}
\end{equation}
Differentiating  (\ref{PCR t}) we obtain the IRF in the form
\begin{equation}\label{H t y}
\begin{aligned}
H&(t, y,I_B,T_b)= -P\left(E\left(t, y,I_B,T_b\right)\right) \Theta\left[I_B-{}\right.\\
&\left.I_{det}(y,T_b,E\left(t, y,I_B,T_b\right))\right]\frac{\partial E\left(t, y,I_B,T_b\right)}{\partial t}\text{.}
\end{aligned}
\end{equation}
The single-valuedness of the solution (\ref{E t})  and negative derivative of latency as a function of energy ensures that the derivative \(\displaystyle{\frac{\partial E(t, y,I_B,T_b)}{\partial t}}\) in expression (\ref{H t y}) is negative. A positive derivative implies the unphysical possibility of the \(PCR\) decreasing with time.

\quad The expression (\ref{H t y}) for the IRF is exact under the assumptions of our model. Defining \(\bar{t} =  \tau_{lat}(y,I_B,T_b,\bar{E})\) and \(\bar{E} = E\left(\bar{t}, y,I_B,T_b\right)\), we transform (\ref{H t y}) to obtain
\begin{equation}\label{H t y 1}
\begin{aligned}
&H(t, y,I_B,T_b)= \\
&-\frac{1}{\sqrt[]{2\pi}\sigma}\exp\left(-\frac{\left[E\left(t, y,I_B,T_b\right) - E\left(\bar{t}, y,I_B,T_b\right)\right]^2}{2\sigma^2}\right)\times \\
&\quad\Theta\left[I_B-I_{det}(y,T_b,E\left(t, y,I_B,T_b\right))\right]\frac{\partial E\left(t, y,I_B,T_b\right)}{\partial t}\text{.}
\end{aligned}
\end{equation}

\quad For small standard deviation \(\sigma\), we may derive the dominant term in the approximation to \(H\left(t, y,I_B,T_b\right)\).  Taking a series expansion up to linear terms in \(\left(t - \bar{t}\right)\), \(E\left(t, y,I_B,T_b\right) - E\left(\bar{t}, y,I_B,T_b\right) = \displaystyle{\left.\frac{\partial E\left(t, y,I_B,T_b\right)}{\partial t}\right|_{t=\bar{t}}\left(t - \bar{t}\right)} = \left(\left.\frac{\partial\tau_{lat}(y,I_B,T_b,E)}{\partial E}\right|_{E=\bar{E}}\right)^{-1} \left(t - \bar{t}\right)\) and substituting the result into (\ref{H t y 1}) yields 
\begin{equation}\label{H t y 2}
\begin{aligned}
H&(t, y,I_B,T_b)= \\
&\frac{1}{\sqrt[]{2\pi}\sigma_j\left(y,I_B,T_b,\bar{E}\right)}\exp\left[-\frac{\left(t - \bar{t}\right)^2}{2{\sigma_j}^2\left(y,I_B,T_b,\bar{E}\right)}\right]\times \\
&\Theta\left[I_B-I_{det}(y,T_b,\bar{E})\right]
\end{aligned}
\end{equation}
where \(\displaystyle{\sigma_j\left(y,I_B,T_b,\bar{E}\right) = \sigma\left|\frac{\partial \tau_{lat}(y,I_B,T_b,\bar{E})}{\partial\bar{E}}\right|}\).  From (\ref{H t y 2}), it follows that the shape of the approximate IRF is a Gaussian. Defining the FWHM of the Gaussian part of the histogram as the timing jitter \(\Upsilon\), we have 
\begin{equation}\label{T Jitter}
\begin{aligned}
\Upsilon&\approx 2.355\sigma_j\left(y,I_B,T_b,\bar{E}\right)\\
&{}\approx 2.355\sigma\left|\frac{\partial \tau_{lat}(y,I_B,T_b,\bar{E})}{\partial\bar{E}}\right|_{\bar{E}=\chi E_\lambda} \text{.}
\end{aligned}
\end{equation}
Averaging the histogram over the transverse coordinate of photon absorption, we derive an approximate result for the mean histogram
\begin{equation}\label{mean T Jitter}
\begin{aligned}
H&(t, I_B,T_b)= \frac{1}{W}\int_{-W/2}^{W/2}dy \frac{1}{\sqrt[]{2\pi}\sigma_j\left(y,I_B,T_b,\bar{E}\right)}\times \\
&\exp\left[-\frac{\left(t - \bar{t}\right)^2}{2{\sigma_j}^2\left(y,I_B,T_b,\bar{E}\right)}\right]\Theta\left[I_B-I_{det}(y,T_b,\bar{E})\right]\text{.}
\end{aligned}
\end{equation}
This averaging results in a distortion of the ideal local Gaussian IRF.  

\quad It is important to emphasize that the detector latency and its derivative 
\(\displaystyle{\left|\frac{\partial \tau_{lat}(y,I_B,T_b,E)}{\partial E}\right|}\) both exhibit a singularity.  They are finite if 
\begin{equation}\label{conditions for singularity}
\begin{aligned}
I_B > I_{det}\left(y, T_b,E\right) \text{ or } E > E_{det}\left(y,I_B ,T_b\right),
\end{aligned}
\end{equation}
where  \(E_{det}\left(y,I_B,T_b\right)\) is the detection (or cutoff) energy.  However, at \(I_B = I_{det}\left(y, T_b,E\right)\) or \(E = E_{det}\left(y,I_B,T_b\right)\), the detector latency and its derivative diverge.  By definition of the detection current or detection energy, \(\tau_{lat}(y,I_B,T_b,E)=\infty\) for either \(I_B \leq I_{det}\left(y, T_b,E\right)\) or \(E \leq E_{det}\left(y, I_B,T_b\right)\) and becomes finite at \(I_B > I_{det}\left(y, T_b,E\right)\) or \(E > E_{det}\left(y, I_B,T_b\right)\).  A stronger singularity will be present in the latency derivative. \(I_{det}\) and \(E_{det}\) are discussed in more detail below.

\quad Another general feature of detector latency as a function of photon energy is its positive curvature. This is intuitively sensible, because the increase in energy due to photon absorption cannot result in an instantaneous break of superconductivity. Although the initial temperature of the quasi-equilibrated distributions of electrons and phonons increases with deposited energy, the latency can only asymptotically approach a small but non-zero value. As a result, the rate of decrease in the latency slows down at higher energies.  This asymptotic behavior combined with the singularity at the detection energy leads to a general positive curvature of the latency as a function of photon energy.  In Section \ref{sec:TDGLModel}, we examine the detector latency by solving the generalized TDGL equations and explicitly demonstrate the accuracy of this statement.  An important consequence of the positive curvature of the latency function can be immediately seen from (\ref{E t}) and (\ref{H t y 1}): for latency time increasing beyond \(\bar{t}\), further reduction of \(E\left(t, y,I_B,T_b\right)\) in the argument of the exponential function in (\ref{H t y 1}) becomes more gradual, resulting in the formation of an extended tail of the IRF.

\quad The presence of the singularity at \(I_{det}\left(y, T_b,E\right)\) results in the general trend of the detector latency \(\tau_{lat}\) and jitter \(\Upsilon\) increasing when the bias current decreases, \(I_B \rightarrow I_{det}\), for all other parameters fixed and assuming noiseless amplification.  If the bias current and other parameters except photon energy are fixed, then both the detector latency and jitter \(\displaystyle{\left(\sim\left|\frac{\partial \tau_{lat}(y,I_B,T_b,E)}{\partial \bar{E}}\right|_{\bar{E}=\chi E_\lambda}\right)}\) monotonically increase as the photon energy decreases to \(E_{det}\left(y, I_b,T_b\right)\), due to the latency being a monotonically decreasing function of energy deposition with positive curvature, as is observed in experiments \cite{korzh_demonstrating_2018}.

\subsection{\label{sec:HBModel}Timing jitter: hotbelt model}

\quad In the hotbelt scenario, \(I_{det}\left(y, T_b,E\right)\) does not depend on \(y\). Solving \(I_B - I_{det}\left(T_b,E^*\right) = 0\) for \(E^*=E_{det}\left(I_B,T_b\right)\) and replacing \(\Theta\left[I_B-I_{det}(T_b,E)\right] \) by \(\Theta\left[E-E_{det}(I_B,T_b)\right] \), we obtain 
\begin{equation}\label{PCR HB}
\begin{aligned}
PCR&(t,I_B,T_b,E_\lambda)=\\
&\int_{E_{det}\left(I_B,T_b\right)}^{E_\lambda}dE P(E)\Theta\left[t - \tau_{lat}\left(I_B,T_b,E\right)\right].
\end{aligned}
\end{equation}
Normalized $PCR$s in the familiar form \cite{kozorezov_fano_2017} are obtained from (\ref{PCR general}-\ref{PCR averaged over y}) by taking the limit \(t \rightarrow \infty \) and \(\Theta\left[t - \tau_{lat}\left(I_B,T_b,E\right)\right]\rightarrow 1\).  This results in 
\begin{equation}\label{PCR HB 1}
\begin{aligned}
PCR(I_B,T_b,\chi,E_\lambda)&=\frac{1}{2}\left[\text{erf}\left(\frac{\chi E_\lambda - E_{det}\left(I_B,T_b\right)}{\sqrt[]{\sigma}}\right)\right.\\
&+\left. \text{erf}\left(\frac{E_\lambda\left(1 - \chi\right)}{\sqrt[]{\sigma}}\right)\right]\text{,}
\end{aligned}
\end{equation}
where \(E_{det}\left(I_B,T_b\right)\) is determined from the energy conservation law.  This conservation law can be written 
\begin{equation}\label{E det HB}
\begin{aligned}
\mathcal{E}_e\left[\left(1 - {I_B}^{2/3}\right)^{1/2}\right]& + \mathcal{E}_{ph}\left[\left(1 - {I_B}^{2/3}\right)^{1/2}\right] ={}\\
\mathcal{E}_e(T_b) &+ \mathcal{E}_{ph}(T_b) + E_{det}\left(I_B,T_b\right)\text{,}
\end{aligned}
\end{equation}
where \(I_B\) is in units of the zero-temperature critical depairing current. The solution of (\ref{E det HB}) determines the threshold for energy deposition into a current carrying superconductor with initial temperature \(T_b\) such that the critical current of the superconductor heated by this energy deposition becomes equal to the bias current \(I_B\),  using the Bardeen relation. Similarly, for a given energy deposition \(E\) the solution of equation (\ref{E det HB}) for current defines the detection current, \(I_{det}\left(E,T_b\right)\).  Substituting (\ref{E det HB}) into (\ref{PCR HB 1}) yields the explicit expression for \(PCR\left(I_B,T_b,\chi,E_\lambda\right)\).  For the hotbelt model, neglecting the $y$-dependence of the detector latency and detection current and simplifying (\ref{H t y 1}), we arrive at 
\begin{equation}\label{main H}
\begin{aligned}
H(t)=-\frac{1}{\sqrt[]{2\pi}\sigma}\exp&\left[-\frac{\left(E\left(t, I_B,T_b\right) - \bar{E}\right)^2}{2{\sigma}^2}\right]\times\\
&\qquad\qquad\qquad\frac{dE\left(t, I_B,T_b\right)}{dt}.
\end{aligned}
\end{equation}
The approximate IRF is similarly obtained from the general result of the hotspot model, acquiring the forms of (\ref{H t y 2}) and (\ref{T Jitter}) where the detector latency, detection current and variance are independent of the transverse coordinate \(y\).

\subsection{\label{sec:SimplfiedHBModel}Understanding intrinsic jitter in SNSPDs using the simplified hotbelt model}
\quad The phenomenological model described above is sufficient to understand and qualitatively interpret the results of recent experiments \cite{korzh_demonstrating_2018} based on the analytical properties of the latency function, its singularities, and its monotonic decrease and positive curvature with increasing energy deposition. Such an analysis becomes especially straightforward for the hotbelt model, where the analytical forms for the \(PCR\) can be used for fitting and for the initial selection of material parameters. However, a quantitatively predictive model requires a more accurate determination of the latency of the detector for an arbitrary deposition energy, which is needed to interpret experimental results of the relative detection latency between pair of photons with different energies \cite{korzh_demonstrating_2018}.  This is sensitive to the detection model, and requires a more sophisticated approach and more extensive numerical analysis. The practical value of the above analysis is that we only need to study the detector latency and fluctuations in order to generate the IRF.

\quad In this sub-section we use the simplified hotbelt model. It is the easiest way to introduce fluctuations and discuss the shape of the photon counting rates as a function of the device parameters \cite{kozorezov_fano_2017}, and it leads to useful qualitative insights into the physics of the detector latency and the intrinsic timing jitter. In Section \ref{sec:TDGLModel}, we will develop a more advanced hotbelt model based on the solution of the 1D generalized time-dependent Ginzburg-Landau (TDGL) equations together with current continuity and thermal balance equations. We will leave the analysis of detector latency in the context of a fully general hotspot model based on the generalized 2D TDGL equations for future work.

\begin{figure}
\includegraphics[width=\linewidth]{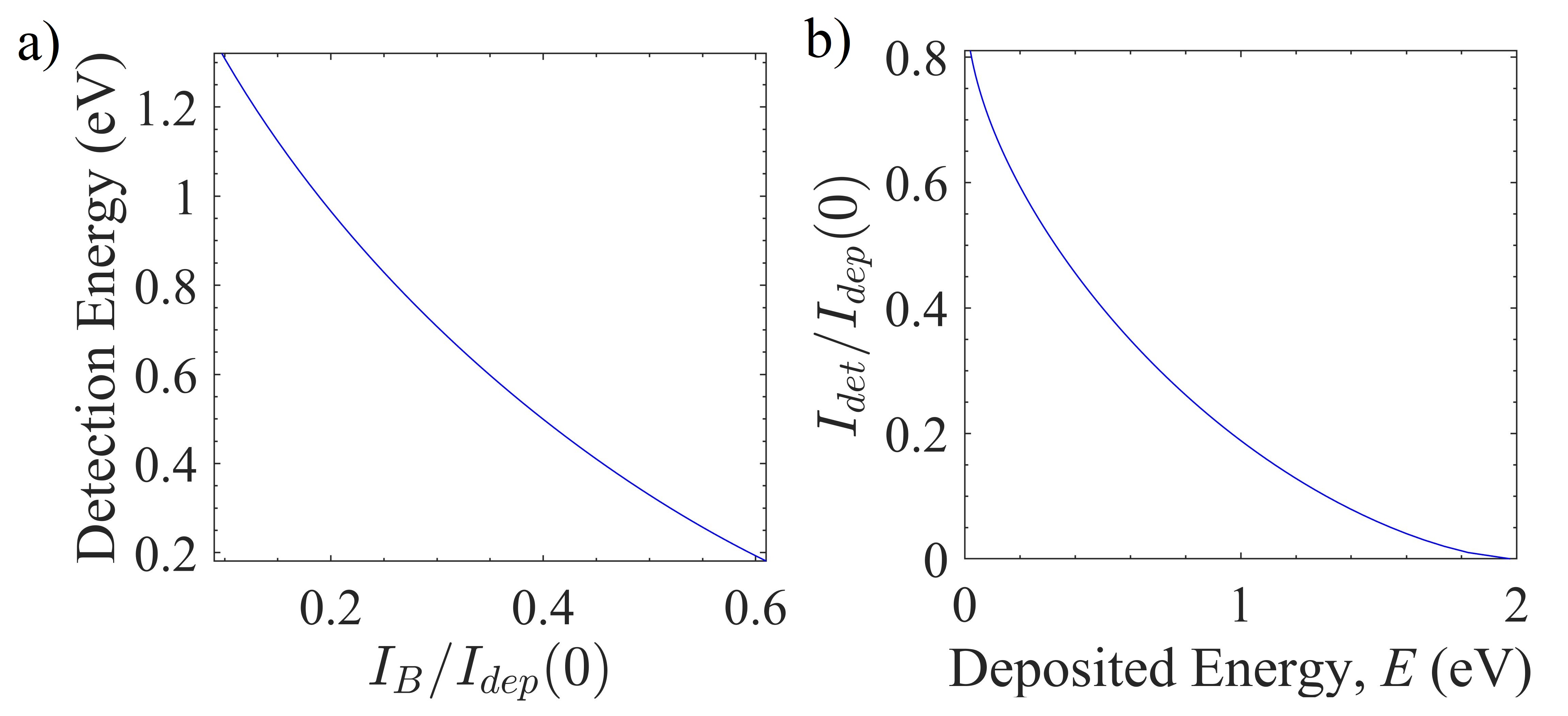} 
\caption{(a) Detection energy \(E_{det}\) vs. bias current \(I_b\) normalized by the depairing critical current at zero temperature \(I_{dep}(0)\). The detection energy at a given \(I_b\) defines the energy deposition \(E\) above which the detector generates a click. (b) Normalized detection current \(I_{det}/I_{dep}(0)\) vs. deposited energy \(E\).  The detection current defines the current above which the detector generates a click for a given amount of deposited energy. Results are calculated using the hotbelt model for bath temperature \(T_b=1~K\) and the material parameters described in the text.}
\label{F HB Cutoff}
\end{figure}

\quad The main results of the simplified hotbelt model are described above by equations (\ref{PCR HB 1}), (\ref{E det HB}) and (\ref{main H}). To illustrate the predictions of the model, we consider the specific case of an NbN SNSPD described in \cite{korzh_demonstrating_2018}: \(W=80 \text{ nm, }d=7 \text{ nm, critical temperature }T_c=8.65\text{ K, }D=0.5 \text{ cm}^2 \text{s}^{-1} \text{, } R_{sq}=587.5\text{ }\Omega/\text{sq} \text{, and } N(0)=1/(2e^2 DR_{sq} d)= 15.2\cdot 10^{21} \text{ cm}^{-3} \text{eV}^{-1}\).  Following \cite{kupriyanov_m._yu._temperature_1980}, we estimate the depairing current at zero temperature for the nanowire using \(I_{dep}(0)=1.491N(0)e\left(\Delta(0)\right)^{3/2} (D/\hbar)^{1/2} Wd \), arriving at \(I_{dep}(0)\approx 26.7\:\mu\text{A, where } \Delta(0)\) is the superconducting gap at zero temperature. The parameter \(\gamma\) is estimated to be 60 based on the acoustic properties of NbN \cite{zou_hexagonal-structured_2015}. Figure~\ref{F HB Cutoff} shows the detection energy and detection current for a square hotbelt \(\left(L = W\right)\) calculated from (\ref{E det HB}). Using the calculations of the detection energy from (\ref{E det HB}) shown in Fig. \ref{F HB Cutoff} (a) and expression (\ref{PCR HB 1}), we can calculate \(PCR\) curves.

\begin{figure} 
\includegraphics[width=0.75\linewidth]{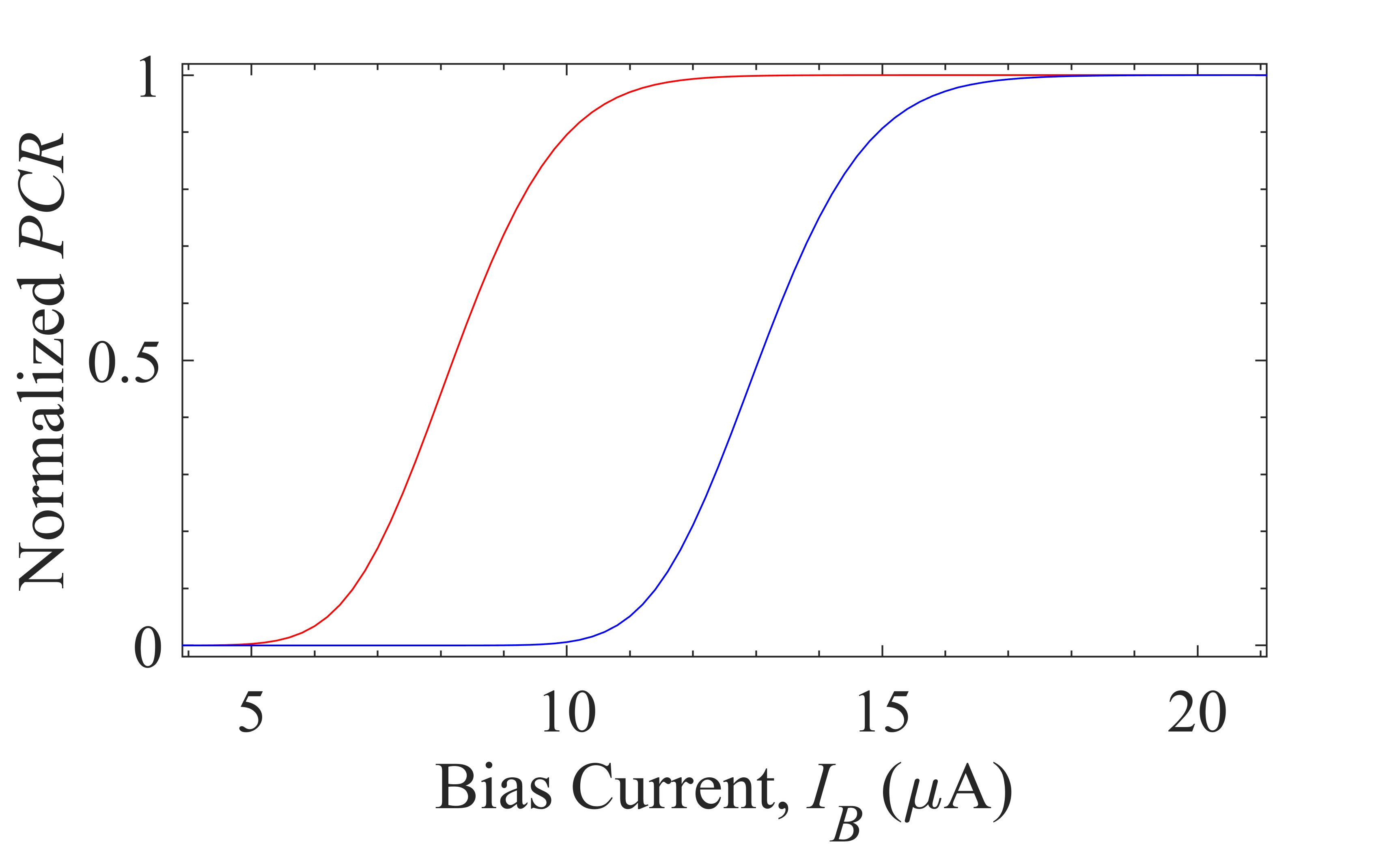}
\caption{Normalized $PCR$ for photon wavelengths of 1550 nm (blue) and 775 nm (red) at \(T_b=1~K\) using the simplified HB model.}
\label{F HB PCR}
\end{figure}

\quad Figure~\ref{F HB PCR} shows the simulated \(PCR\) curves for 1550 and 775 nm photons at \(T_b=1\) K, which closely follow the data presented in \cite{korzh_demonstrating_2018} if we use \(\chi=0.43\). This value can be justified as follows: energy loss from the quasi-equilibrated quasiparticles and phonons making up the hotbelt occurs due to the out-diffusion of quasiparticles and the escape of phonons. Cooling of the hotbelt and the order parameter suppression takes a substantial part of the detector latency interval, terminating with the SNSPD entering the resistive state. During this stage, the cooling rate is determined by a combination of the phonon escape rate \(\tau_{esc}^{-1}\) and the thermal conductivity, both of which are independent of the bias current and the initial state determined by the photon energy. Therefore, the energy loss from electrons and phonons is determined by the ratio \(\tau_{lat}\left(I_B,T_b,\bar{E},\right)/\text{min}\left\lbrace \tau_{esc},L^2/4D \right\rbrace\); the larger this ratio is, the more energy is lost.  The energy of the quasiparticles and phonons in the hotbelt is a monotonically decreasing function of this ratio. When \(\tau_{lat}\left(I_B,T_b,\bar{E},\right)/\text{min}\left\lbrace \tau_{esc},L^2/4D \right\rbrace \ll 1\), there is no loss of photon energy. In the opposite case, i.e. \(\tau_{lat}\left(I_B,T_b,\bar{E},\right)/\text{min}\left\lbrace \tau_{esc},L^2/4D \right\rbrace \gg 1\), the loss is substantial.  Approximately half the photon energy is lost when \(b\tau_{lat}\left(I_B,T_b,\bar{E},\right)/\text{min}\left\lbrace \tau_{esc},L^2/4D \right\rbrace = 1\), where \(b \leq 1\) is a numerical factor of order unity accounting for the fact that only the first part of the latency interval is dissipationless until the superconducting current flow becomes unstable.  In typical situations, \(\tau_{esc}=4d/\eta c\) \cite{kaplan_acoustic_1979} where \(d\geq 5 \text{ nm}\), the phonon transmission coefficient through the escape interface \(\eta \sim 0.3\), and mean sound velocity \(c \sim 5 \cdot 10^5 \text{ cm/s}\), we find \(\tau_{esc} \sim 15 \text{ ps}\) and \(L^2/4D \sim 30 \text{ ps} \) for \(L=80 \text{ nm}\), and the detector latency is either shorter or close to  \(\text{min}\left\lbrace \tau_{esc},L^2/4D \right\rbrace\).  Under these conditions we expect the middle of the interval to be a good representative value and \(\chi=1/2\left(1 + 4\pi^2/5\gamma\right)\), where the fraction \(1/\left(1 + 4\pi^2/5\gamma\right)\) accounts for the energy in the electronic system.

\begin{figure}
\includegraphics[width=0.75\linewidth]{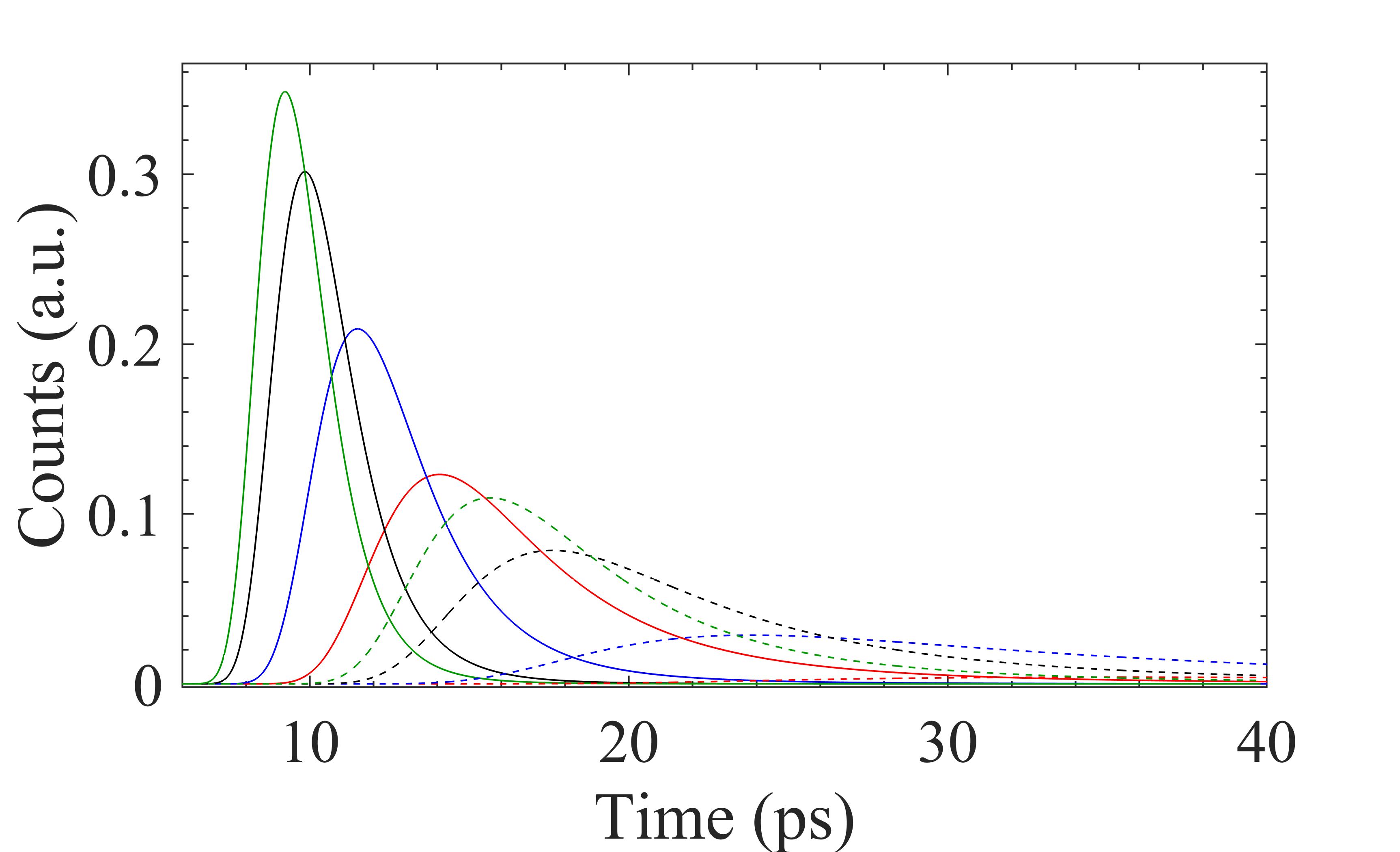}
\caption{Instrument response functions for 1550 nm (dashed) and 775 (solid) nm photons at bias currents of \(12\: \mu\)A (red), \(14\: \mu\)A (blue), \(16\: \mu\)A (black), and \(17\: \mu\)A (green). The critical depairing current at zero temperature is \(26.7\: \mu\)A.}
\label{F Histograms}
\end{figure}

\begin{figure}
\includegraphics[width=\linewidth]{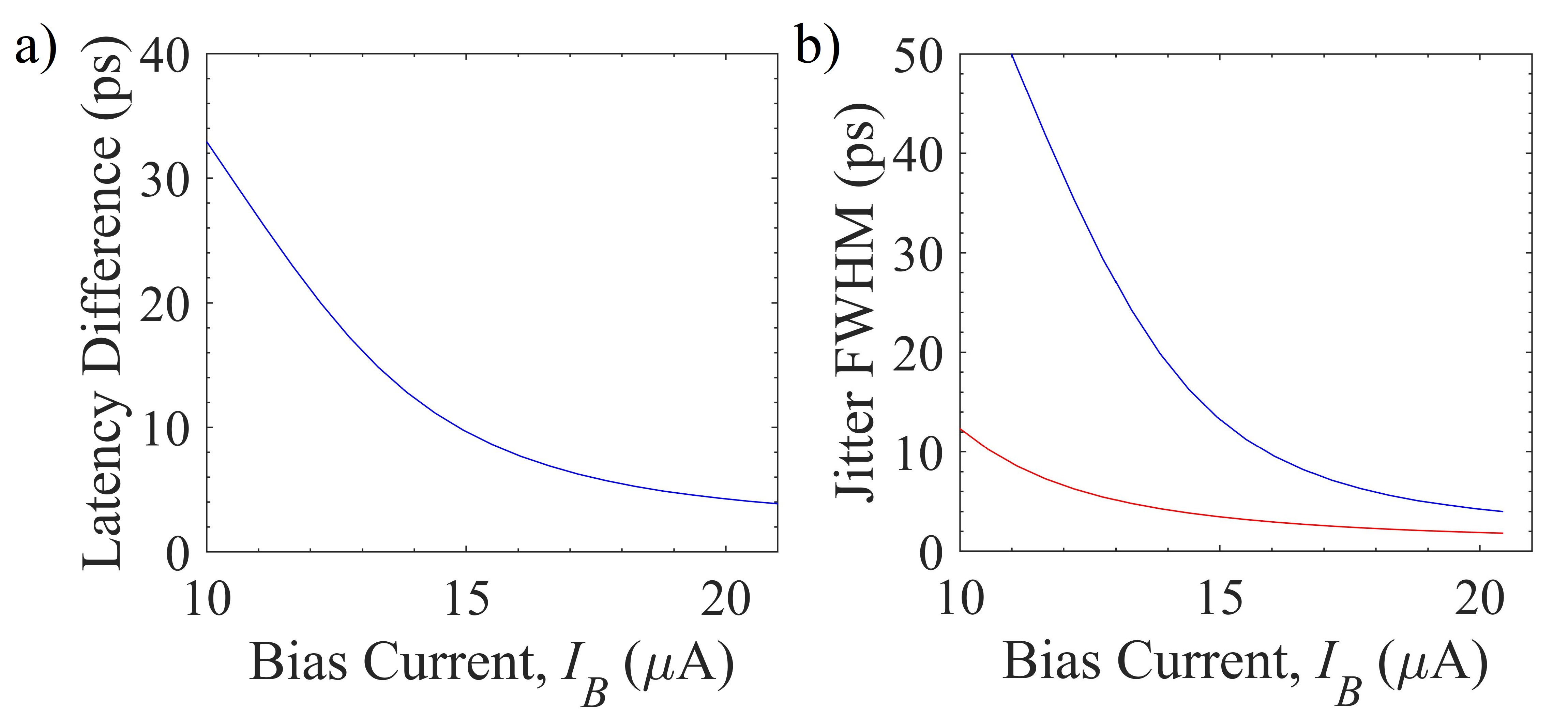}
\caption{(a) Relative latency vs bias current for 1550/775 nm photons; (b) Timing jitter vs bias current for 775 nm (red) and 1550 nm (blue) incident photons.}
\label{F HB Latency}
\end{figure}

\quad To model the jitter, we need to use the approximation for the detector latency 
\(\tau_{lat}\left(I_B,T_b,E\right)\), 
which must be found from more advanced detection models as explained in the previous sections. Here we exploit its singular character at
\(E_{det}\left(I_B,T_b\right)\)
or at 
\(I_{det}\left(E,T_b\right)\)
and its monotonic decrease as
\(E \rightarrow \infty\),
choosing the forms
\begin{equation}\label{appr tau lat E}
\begin{aligned}
\tau_{lat}&\left(I_B,T_b,E\right) = \\
&\tau_{lat}\left(I_B,T_b,E_0\right)\left(\frac{E_0-E_{det}\left(I_B,T_b\right)}{E-E_{det}\left(I_B,T_b\right)}\right)^\alpha \\
&{}
\end{aligned}
\end{equation}
and
\begin{equation}\label{appr tau lat I}
\begin{aligned}
\tau_{lat}&\left(I_B,T_b,E\right) = \\
&\tau_{lat}\left(I_{sw},T_b,E\right)\left(\frac{I_{sw}-I_{det}\left(E,T_b\right)}{I_B-I_{det}\left(E,T_b\right)}\right)^\alpha \text{.}
\end{aligned}
\end{equation}
Here
\(E_0 = 0.8\text{ eV}\) is the reference energy of a 1550~nm photon and \(I_{sw}\) is the switching current of the SNSPD.  Eq. (\ref{appr tau lat I}) reflects the singularity in the latency when the current approaches the detection current for a deposited energy \(E\). The exponents in (\ref{appr tau lat E}) and (\ref{appr tau lat I}) are the same. When 
\(E \rightarrow E_{det}\left(I_B, T_b\right)\text{,}\)
 we have 
\(I_B - I_{det}\left(E, T_b\right) \approx \frac{\partial I_{det}}{\partial E} \left|_{E_{det}\left(I_B,T_b\right)}\left(E - E_{det}\left(I_B, T_b\right)\right)\right. \)
neglecting higher order terms. The derivative $\displaystyle{\frac{\partial I_{det}}{\partial E}}$ is continuous (see Fig. \ref{F HB Cutoff} (b)) and (\ref{appr tau lat I}) exhibits the same singularity as (\ref{appr tau lat E}). Combining (\ref{appr tau lat E}) and (\ref{appr tau lat I}), we obtain
\begin{equation}\label{comb appr tau lat}
\begin{aligned}
&\tau_{lat}\left(I_B,T_b,E\right) = \tau_{lat}\left(I_{sw},T_b,E_0\right)\times\\
&\left(\frac{I_{sw}-I_{det}\left(E_0,T_b\right)}{I_B-I_{det}\left(E_0,T_b\right)}\right)^\alpha \left(\frac{E_0-E_{det}\left(I_B,T_b\right)}{E-E_{det}\left(I_B,T_b\right)}\right)^\alpha .
\end{aligned}
\end{equation}
For simplicity, we assume that the exponential \(\alpha\) does not depend explicitly on \(I_B\) and \(T_b\), and also that
\(\text{lim}_{E\rightarrow \infty} \tau_{lat}\left(I_B, T_b, E\right) = 0 \), i.e. that the detector responds instantaneously to an infinite energy deposition. Here, 
\(\tau_{lat}\left(I_{sw}, T_b, E_0\right) \)
is the latency at some reference energy, in this case corresponding to \(\lambda = 1550 \text{ nm}\), and at $I_B=I_{sw}$. Solving (\ref{E t}) we obtain
\begin{equation}\label{E(t)}
\begin{aligned}
&E\left(t\right) = E_{det}\left(I_B,T_b\right) + \left(E_0 - E_{det}\left(I_B,T_b\right)\right)\times\\
&\left(\frac{I_{sw} - I_{det}\left(E_0,T_b\right)}{I_B - I_{det}\left(E_0,T_b\right)}\right)\left(\frac{\tau_{lat}\left(I_{sw},T_b,E_0\right)}{t}\right)^{1/\alpha} \text{.}
\end{aligned}
\end{equation}
Substituting this result into (\ref{main H}) we finally obtain
\begin{widetext}
\begin{equation}\label{Simple HB H}
\begin{aligned}
H\left(t\right) ={}& \frac{1}{\sqrt[]{2\pi}\sigma} \exp \left\lbrace - \frac{\left[E_{det}\left(I_B,T_b\right) + \left(E_0 - E_{det}\left(I_B,T_b\right)\right)\left(\frac{I_{sw} - I_{det}\left(E_0,T_b\right)}{I_B - I_{det}\left(E_0,T_b\right)}\right)\left(\frac{\tau_{lat}\left(I_{sw},T_b,E_0\right)}{t}\right)^{1/\alpha} - \chi E_\lambda \right]^2 }{2 \sigma^2} \right\rbrace \times \\
&\frac{\left(E_0 - E_{det}\left(I_B,T_b\right)\right)}{\alpha t}\left(\frac{I_{sw} - I_{det}\left(E_0,T_b\right)}{I_B - I_{det}\left(E_0,T_b\right)}\right)\left(\frac{\tau_{lat}\left(I_{sw},T_b,E_0\right)}{t}\right)^{1/\alpha} \text{.}
\end{aligned}     
\end{equation}
\end{widetext}
The expression (\ref{Simple HB H}) can now be used for the analysis of latency difference and IRF. For illustration, we enter the parameters for an 80~nm-wide NbN SNSPD with \(I_{sw} = 21~\mu\)A, \(\tau_{lat}\left(I_{sw},T_b,E_0\right) = 7\) ps, and \(\alpha = 0.6 \).  Figure~\ref{F Histograms} shows the calculated IRFs for several different currents for 1550 and 775 nm photons at \(T_b = 1 \text{ K}\).
Figure~\ref{F HB Latency} (a) shows the latency difference for pairs of 1550 and 775 nm photons extracted from the simulations shown in Fig. \ref{F Histograms}. Figure~ \ref{F HB Latency} (b) presents the results for the FWHM of the IRF calculated with the use of (\ref{comb appr tau lat}) - (\ref{Simple HB H}).

\quad The simulated curves in Figs. \ref{F HB PCR}-\ref{F HB Latency} are qualitatively consistent with the results of recent experiments \cite{korzh_demonstrating_2018}, providing support to the conclusion that intrinsic jitter was observed in this experiment. It is remarkable that incorporation of the general features of detector latency alone, namely singularities at the detection current and detection energy, together with monotonic positive curvature variation with energy deposition, is sufficient to qualitatively reproduce the intrinsic timing jitter. The model succeeds in reproducing all the primary experimental features: strong dependence of the jitter on bias current, similar functional behavior of the latency difference and timing jitter, and transformation of the IRF distributions with the bias current.  In addition, it allows for satisfactory fitting of the observed $PCR$ vs bias current curves for different wavelengths. Even the magnitudes of the observed latency difference, timing jitter, and counting rates fall in the same range with reasonable accuracy.  Nonetheless, the results of the simulations are sensitive to the exact functional dependence of \(\tau_{lat}\left(I_{B},T_b,E\right)\), the magnitudes of \(\tau_{lat}\left(I_{sw},T_b,E_0\right)\) and  \(\alpha\), and the possibility to model detector latency in the form of (\ref{comb appr tau lat}). The primary disadvantage of the simple hotbelt theory is its inherent incapability, because of its phenomenological character, to derive the detector latency from the basic principles of non-equilibrium superconductivity. 

\section{\label{sec:TDGLModel}Time dependent Ginzburg-Landau model of detector latency}

\quad The simple phenomenological hotbelt model of latency and timing jitter described above is insufficient to provide a quantitatively predictive picture of the latency characteristics of SNSPDs.  To rigorously describe the detector latency, we must use a more advanced technique which includes the simulation of the evolving non-equilibrium superconducting state in combination with numerical modeling.  To analyze the evolution of the non-equilibrium state caused by photon absorption in a superconducting nanowire and understand the characteristics of detector latency in SNSPDs, we use the generalized time-dependent Ginzburg-Landau model.  

\subsection{\label{sec:GenTDGL}Generalized TDGL formulation}
Different microscopic models can be compared and validated by predicting the detector latency and comparing it to experiments.  At present, the most advanced microscopic model of an SNSPD uses electrothermal equations coupled to TDGL equations describing the superconducting order parameter in a two-dimensional system \cite{vodolazov_single-photon_2017}.  In this formulation, the energy balance in the electron and phonon systems take the form

\begin{equation}\label{e balance}
\begin{aligned}
\frac{d}{dt}&\left(\frac{\pi^2 {k_B}^2 N(0) {T_e}^2}{3} - E_c \mathcal{E}_s \left(T_e,\left|\Delta\right|\right)  \right) = \\
&\nabla k_s \nabla T_e - \frac{2 {\pi}^2 {k_B}^2 N(0)}{15 \tau_{ep}\left(T_c\right)}\frac{{T_e}^5 - {T_{ph}}^5}{{T_c}^3} + \vec{j}\cdot\vec{E}
\end{aligned}     
\end{equation}

\begin{equation}\label{Ph balance}
\begin{aligned}
\frac{d {T_{ph}}^4}{dt} = \frac{\gamma}{2 {\pi}^2 \tau_{ep}\left(T_c\right)} \frac{{T_e}^5 - {T_{ph}}^5}{T_c} - \frac{{T_{ph}}^4 - {T_b}^4}{\tau_{esc}} \text{,}
\end{aligned}     
\end{equation}
where \(k_s\) is the thermal conductivity of the electron system in the superconducting state 

\begin{equation}\label{Bardeen}
\begin{aligned}
k_s = \frac{2D\pi^2 {k_B}^2 N(0) T_e}{3} \left(1 - \frac{6}{\pi^2}\int_0^{\left|\Delta\right| /  k_B T_e} \frac{x^2 e^x dx}{\left(e^x + 1\right)^2 } \right)
\end{aligned}     
\end{equation}
and Joule heating is included in the electron energy equation as the dot product of the total current density 
\(\vec{j}\) and electric field \(\vec{E}\).  The energy gain of the electron system due to transitioning to the superconducting state \(\mathcal{E}_s\) is given by 

\begin{equation}\label{ES}
\begin{aligned}
\mathcal{E}_s &{}= \int_0^{\left|\Delta\right|/k_B T_e} \tilde{\epsilon} n_{\tilde{\epsilon}} d \tilde{\epsilon} \\
&{}- \int_{\left|\Delta\right| / k_B T_e}^{\infty} \tilde{\epsilon} \left(\frac{\tilde{\epsilon}}{\sqrt[]{\tilde{\epsilon}^2 - \left(\frac{\left|\Delta\right|}{k_B T_e} \right)^2}} - 1 \right) n_{\tilde{\epsilon}} d \tilde{\epsilon} \\
&{}+ \left(\frac{\left|\Delta\right|}{2 k_B T_c}\right)^2 \left(\frac{1}{2} + \text{ln}\left(\frac{\Delta_0}{\left|\Delta\right|}\right)\right).
\end{aligned}     
\end{equation}
The superconducting order parameter at zero temperature is given by 
\(\Delta_0 = 1.764\: k_B T_c \), and \(n_{\tilde{\epsilon}}\) in (\ref{ES}) is the Fermi distribution.  The total current density \(\vec{j} = \vec{j}_n + \vec{j}_s\) is the sum of the normal current
\begin{equation}\label{normal j}
\begin{aligned}
\vec{j}_n = -\sigma_n \nabla \varphi
\end{aligned}     
\end{equation}
and the supercurrent, for which we use a close approximation to the general Usadel result
\begin{equation}\label{Usadel j}
\begin{aligned}
\vec{j}_s = \vec{j}_s^{Us} \simeq \frac{\pi \sigma_n}{2 e \hbar}\left|\Delta\right|\tanh\left(\frac{\left|\Delta\right|}{2 k_B T_e}\right) \text{,}
\end{aligned}     
\end{equation}
where \(\varphi\) is the electrostatic potential and \(\sigma_n\) is the conductivity in the normal state \cite{vodolazov_single-photon_2017}.  The order parameter \(\Delta = \left|\Delta\right|e^{i \phi} \) is described by the dynamics of the TDGL equations.  It is commonly known that the standard TDGL equations are only valid when the temperature is close to \(T_c\) and deviations from equilibrium are small.  The equations also only apply in the limiting case of a gapless superconductor satisfying \(\left|\Delta\right| \tau_{s\_mag} \ll \hbar \), where \(\tau_{s\_mag}\) is the magnetic scattering time.  In the following work, we use a more general version of the TDGL equations with less stringent requirements \cite{watts-tobin_nonequilibrium_1981,kopnin_theory_2001}.  For a dirty superconductor with strong impurity scattering, the assumptions of slow variations in time and space are no different from those of the standard TDGL.  Derivation of the generalized TDGL equations does not require the strong limitations necessary for gapless superconductivity.  Correspondingly, the generalized TDGL equations better suit simulating the suppression of the gap over extended intervals of time when both the superconducting order parameter and the energy gap remain finite.  This is exactly the case for the order parameter dynamics following the absorption of a photon.  The precise details of the order parameter evolution are required for reliable description of temporal properties of SNSPDs such as the latency and timing jitter.  

\quad The generalized TDGL equations can be written as the time dependent partial differential equations
\begin{equation}\label{G TDGL}
\begin{aligned}
\frac{\pi \hbar}{8 k_B T_c}&\left(\varrho\left(T_e\right)\frac{\partial}{\partial t}\left|\Delta\right| + \frac{i\left|\Delta\right|}{\varrho\left(T_e\right)}\frac{\partial}{\partial t} \phi + \frac{2 i e \left|\Delta\right|}{\varrho\left(T_e\right) \hbar}\varphi\right) = \\
&\xi_{mod}\left(T_e\right)^2\left(\nabla + i\left(\nabla\phi - \frac{2e}{\hbar c} A\right)\right)^2 \left|\Delta\right| \\
&+ \left(1 - \frac{T_e}{T_c} - \frac{\left|\Delta\right|^2}{\Delta_{mod}^2\left(T_e\right)}\right)\left|\Delta\right| \\
&+ i\frac{\left(\nabla\cdot\vec{j}_s^{Us} - \nabla\cdot\vec{j}_s^{GL}\right)}{\left|\Delta\right|}\frac{\hbar e D}{\sigma_n\sqrt[]{2}\:\sqrt[]{1 + T_e/T_c}}
\end{aligned}     
\end{equation}
where the parameter
\(\varrho\left(T_e\right) = \sqrt[]{1 + \left|\Delta\right|^2\tau_{sc}\left(T_e\right)^2/\hbar^2} \)
modifies the rates of phase and magnitude evolution of the TDGL equations.  The terms 
\({\xi_{mod}\left(T_e\right)}^2 = \pi\hbar D/\left(4\:\sqrt[]{2}k_B T_c\:\sqrt[]{1 + T_e/T_c}\right)\)
and
\(\Delta_{mod}^2\left(T_e\right) = \left(\Delta_0\tanh\left(1.74\:\sqrt[]{T_c/T_c - 1}\right)\right)^2 / \left(1 - T_e/T_c\right)\)
 are modified as suggested in \cite{vodolazov_single-photon_2017} in order to closely match the correct temperature dependences well below \(T_c\).  The generalized TDGL equations break the symmetry between the evolving phase and magnitude of the order parameter because the relaxation  of the order parameter magnitude is controlled by a different process than the relaxation of its phase.  Consequently, they cannot be written in the standard TDGL form.  Inelastic scattering incorporates both electron-electron and electron-phonon interactions according to \(\tau_{sc}\left(T_e\right) = 1/\left(1/\tau_{ee}\left(T_e\right) + 1/\tau_{ep}\left(T_e\right)\right) \)
where 
\(\tau_{ee}\left(T_e\right)\)
and
\(\tau_{ep}\left(T_e\right)\)
are the electron-electron and electron-phonon inelastic scattering times respectively.  The temperature dependence of these scattering rates is defined by 
\(\tau_{ee}\left(T_e\right) = \tau_{ee}\left(T_c\right)T_c/T_e\)
and 
\(\tau_{ep}\left(T_e\right) = \tau_{ep}\left(T_c\right)\left(T_c/T_e\right)^3\). 
The final term of (\ref{G TDGL}) enforces the conservation of the Usadel supercurrent in the stationary state following \cite{vodolazov_single-photon_2017}.  Conservation of total current density, 
\(\nabla\cdot\vec{j} = 0 \)
, is enforced with an additional equation
\begin{equation}\label{current conservation}
\begin{aligned}
\sigma_n{\nabla}^2\varphi = \nabla\cdot\vec{j}_s.
\end{aligned}     
\end{equation}

\quad The boundary conditions at the ends of the nanowire are defined by 
\(T_e = T_c\), \(T_{ph} = T_b\), \(\left|\Delta\right| = 0\), \(\vec{j}_s\left|_n\right.=0\), and \(\vec{j}_n\left|_n\right.=I/Wd\).  The introduction of current to the simulation domain through \(\vec{j}_n\) is numerically easier than through \(\vec{j}_s\), but leads to runaway Joule heating of the superconductor.  This is solved by limiting Joule heating to the domain of interest in the center of the nanowire and enhancing cooling through electron-phonon coupling near the nanowire edges. Demagnetization effects are negligible in thin and narrow nanowires, so the vector potential \(A\) is neglected in the absence of a magnetic field.  The system of equations described by (\ref{e balance},\ref{Ph balance},\ref{G TDGL},\ref{current conservation}) is solved numerically in one dimension.  The system is first allowed to evolve to a stationary state configuration for a fixed bias current.  Once the system has stabilized, a fixed amount of energy is added to the electron and phonon systems in a hotbelt of length \(L\) such that \(T_e = T_{ph} \neq T_b \).  This excitation serves as the initial conditions for the system, and the resulting evolution models the response of the nanowire to an excitation of known energy.

\quad For simplicity, we consider a constant bias current in our model.  In electrothermal simulations, SNSPDs are typically modeled as a variable resistor in series with an inductor representing the kinetic inductance of the superconductor \cite{yang_modeling_2007}.  This circuit is then coupled to an external readout circuit used to record the electrical pulse generated during a detection event.  In an SNSPD, current is diverted from the nanowire to the readout path once a significant potential forms across the hotspot.  The electric potential rises quickly after the first phase slip or vortex crossing occurs, but remains small until that point.  Therefore, the current will not be diverted until multiple phase slips have occurred and the growth of the normal domain is dominated by the thermal balance of Joule heating and cooling through diffusion or phonon coupling to the substrate.  In the 1D hotbelt model at bias currents high enough for photodetection, the first phase slip event leads to runaway Joule heating.  This further justifies the use of a constant bias current for determining the relative latency characteristics within this simplified model because the latency is primarily determined by the time for the first phase slip to occur.  This behavior is unlike the case of the 2D model where vortices can cross the nanowire without nucleating a normal domain \cite{vodolazov_single-photon_2017}, but we leave the analysis of the generalized TDGL equations within a 2D framework for the future.  The choice of constant bias current is further supported by experimental evidence that photon energy does not affect the shape of the electrical signal generated during a photodetection event \cite{korzh_demonstrating_2018}.

\subsection{\label{sec:TDGLLatency}Detector latency}
\begin{figure}
\includegraphics[width=\linewidth]{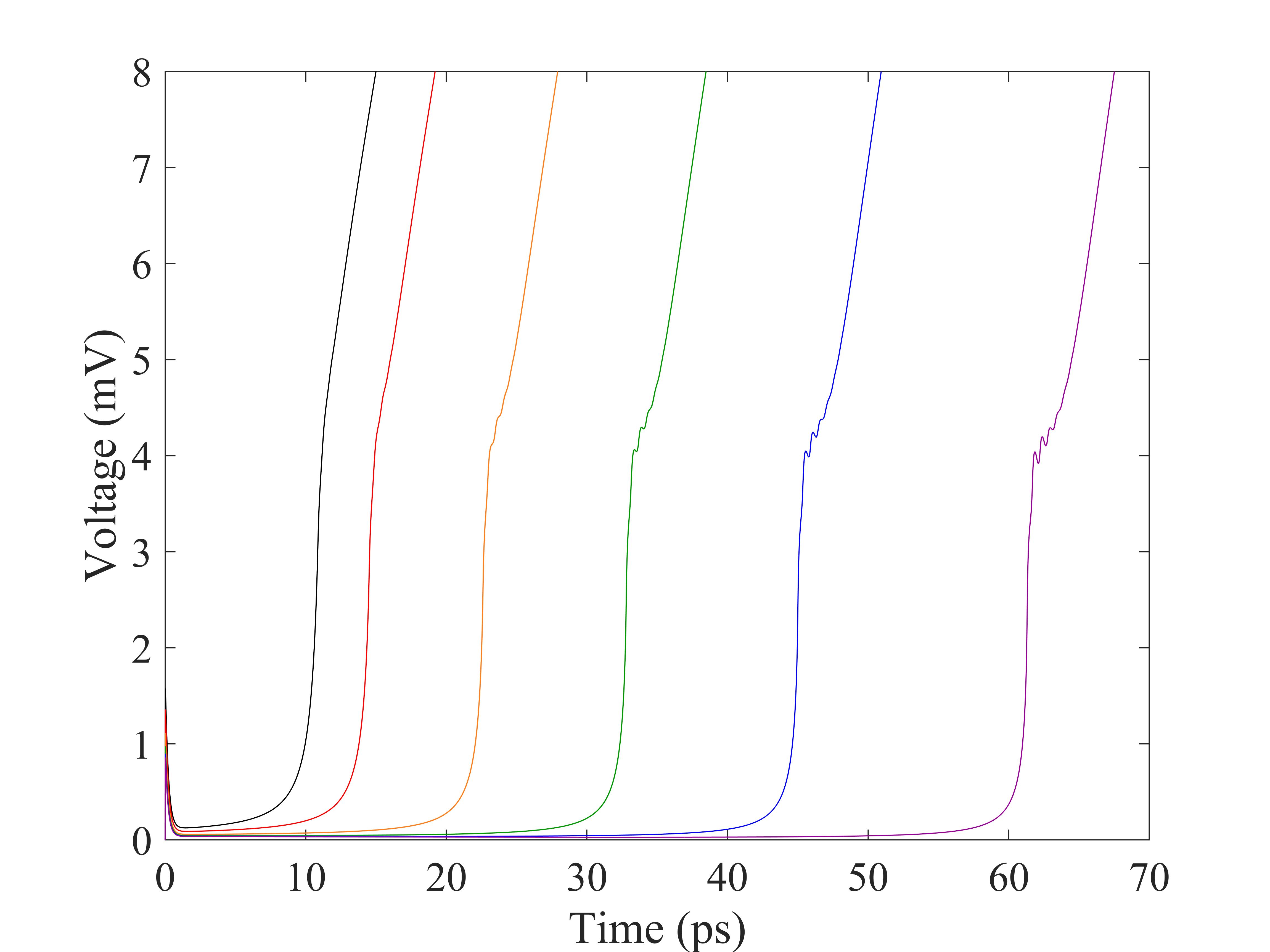}
\caption{Voltage curves for a bias current of 12.5 \(\mu\)A and \(\tau_{ee}\left(T_c\right) \) of 5 ps for energy depositions of 0.8 eV (black), 0.64 eV (red), 0.48 eV (orange), 0.36 eV (green), 0.34 eV (blue), and 0.33 eV (violet).  The voltage is initially greater than zero due to the temperature dependence of the definition of the Usadel supercurrent (\ref{Usadel j}) and the instantaneous increase in \(T_e\) due to the photon energy deposition.  The oscillations due to  individual phase slips are small compared to the voltage oscillations due to vortex crossing in \cite{vodolazov_single-photon_2017} because the order parameter is suppressed over the entire hotbelt length rather than a distance on the order of \(\xi_c\).
}
\label{F Vtrace}
\end{figure}

Calculations are performed for the same 80 nm wide nanowire parameters listed in Section \ref{sec:SimplfiedHBModel} at a substrate temperature \(T_b\)= 2 K.  This choice allows for easier numerical computation compared to 1 K and is expected to be representative of experimental results at 1 K based on the temperature dependence measurements of \cite{korzh_demonstrating_2018}.  The parameter defining the electron-electron inelastic scattering time \(\tau_{ee}\left(T_c\right) \) is taken as an unknown parameter, while \(\tau_{esc}\) is 20 ps and \(\tau_{ep}\left(T_c\right) \) is 24.7 ps based on measurements of \(\tau_{ep}\left(T_c\right) \) of 16 ps for NbN with a \(T_c\) of 10 K scaled by an inverse cubic power law \cite{vodolazov_single-photon_2017,semenov_analysis_1995}.  An initial hotbelt length of $L=80$ nm is used for all simulations.  Once the temperature of the hotbelt is elevated to the initial temperature governed by the photon energy, the system evolves until either a normal domain forms or the photon-induced excitation relaxes back to the superconducting state. Simulated voltage traces are shown in Fig. \ref{F Vtrace} for a bias current of 12.5 \(\mu\)A and \(\tau_{ee}\left(T_c\right) \) of 5 ps under different values of energy deposition.  Lower energy depositions that are above the detection energy lead to longer waiting times before a voltage pulse appears.

\begin{figure}
\includegraphics[width=\linewidth]{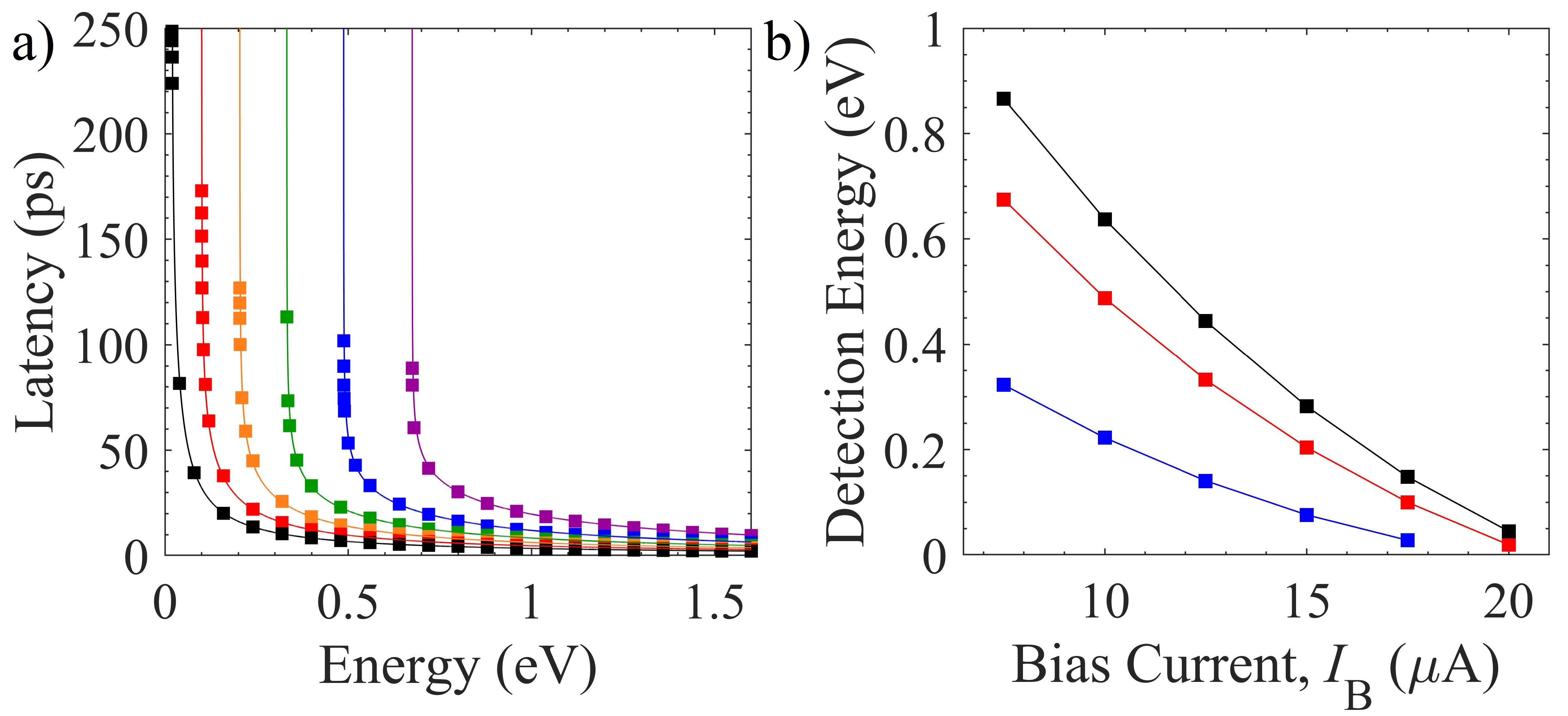}
\caption{(a) Latency vs. energy simulation results for \(\tau_{ee} \left(T_c\right)\) of 5 ps for bias currents of 7.5 \(\mu\)A (violet), 10 \(\mu\)A (blue), 12.5 \(\mu\)A (green), 15 \(\mu\)A (orange), 17.5 \(\mu\)A (red), and 20 \(\mu\)A (black).  The associated fits (lines) are shown for each bias current.  (b) Detection energy vs. bias current for \(\tau_{ee}\left(T_c\right) \) of 0 ps (blue), 5 ps (red), and 10 ps (black).
}
\label{F Latency and Cutoff}
\end{figure}

By selecting a fixed voltage threshold, characteristic latency curves are determined as a function of energy deposited in the superconducting system.  In typical experiments, this threshold is chosen to maximize the slew rate of the rising edge of the pulse in order to minimize the effect of electrical noise-induced timing jitter \cite{korzh_demonstrating_2018}.  In our model, the choice is less crucial because a given bias current leads to normal domain growth at a fixed rate once the first phase slips nucleate this normal domain.  As a result, the slopes of the voltage curves are approximately the same and a shift in the threshold leads to a latency offset, but no change in the relative latency characteristics for different amounts of incident energy.  The latency results are fitted with curves of the form 
\begin{equation}\label{TDGL model differential latency}
\begin{aligned}
\tau_{lat}\left(I_B, \right.&\left.T_b, E\right) = \left(\tau_{lat}\left(I_B,T_b,E_0\right) - \tau_{lat}\left(I_B,T_b,\infty\right)\right)\times \\
&{\left(\frac{E_0 - E_{det}\left(I_B,T_b\right)}{E - E_{det}\left(I_B,T_b\right)} \right)}^{\alpha\left(I_B,T_b\right)}{\left(\frac{E_0}{E}\right)}^{s\left(I_B,T_b\right)}\\
&+ \tau_{lat}\left(I_B,T_b,\infty \right)
\end{aligned}     
\end{equation}
which provide reasonable fits at all energies above \(E_{det}\left(I_B,T_b\right)\) as shown in Fig. \ref{F Latency and Cutoff} (a).  The hotbelt TDGL model predicts a monotonically decreasing positive curvature latency as photon energy increases, as predicted by the phenomenological arguments of Section \ref{sec:SimplfiedHBModel}.  The expression (\ref{TDGL model differential latency}) contains the term describing the singularity of the latency at the detection energy, but note that the form (\ref{TDGL model differential latency}) is more general than the simplified approximation used in the Section \ref{sec:SimplfiedHBModel}, formulas (\ref{appr tau lat E})-(\ref{comb appr tau lat}). It contains the extra factor 
\({\left(\frac{E_0}{E}\right)}^{s\left(I_B,T_b\right)}\) and explicit bias and temperature dependence of the exponents \(\alpha\left(I_B,T_b\right)\) and \(s\left(I_B,T_b\right)\).  The extra factor could have been added to expressions (\ref{appr tau lat E})-(\ref{comb appr tau lat}) on the grounds that the product of a continuous function of energy and the latency must exhibit a singularity at the detection energy.  However, we preferred the simpler approximations commensurate with the basic character of the simple hotbelt model of Section \ref{sec:SimplfiedHBModel}. 

\quad It is worth noticing that the detection energy curve for the simple HB model of Section \ref{sec:SimplfiedHBModel}, which was derived from (5) and shown in the Fig. \ref{F HB Cutoff} (a), falls between the red and black curves of Fig. \ref{F Latency and Cutoff} (b).

\subsection{\label{sec:TDGLIRF}Instrument response function}
\begin{figure}
\includegraphics[width=\linewidth]{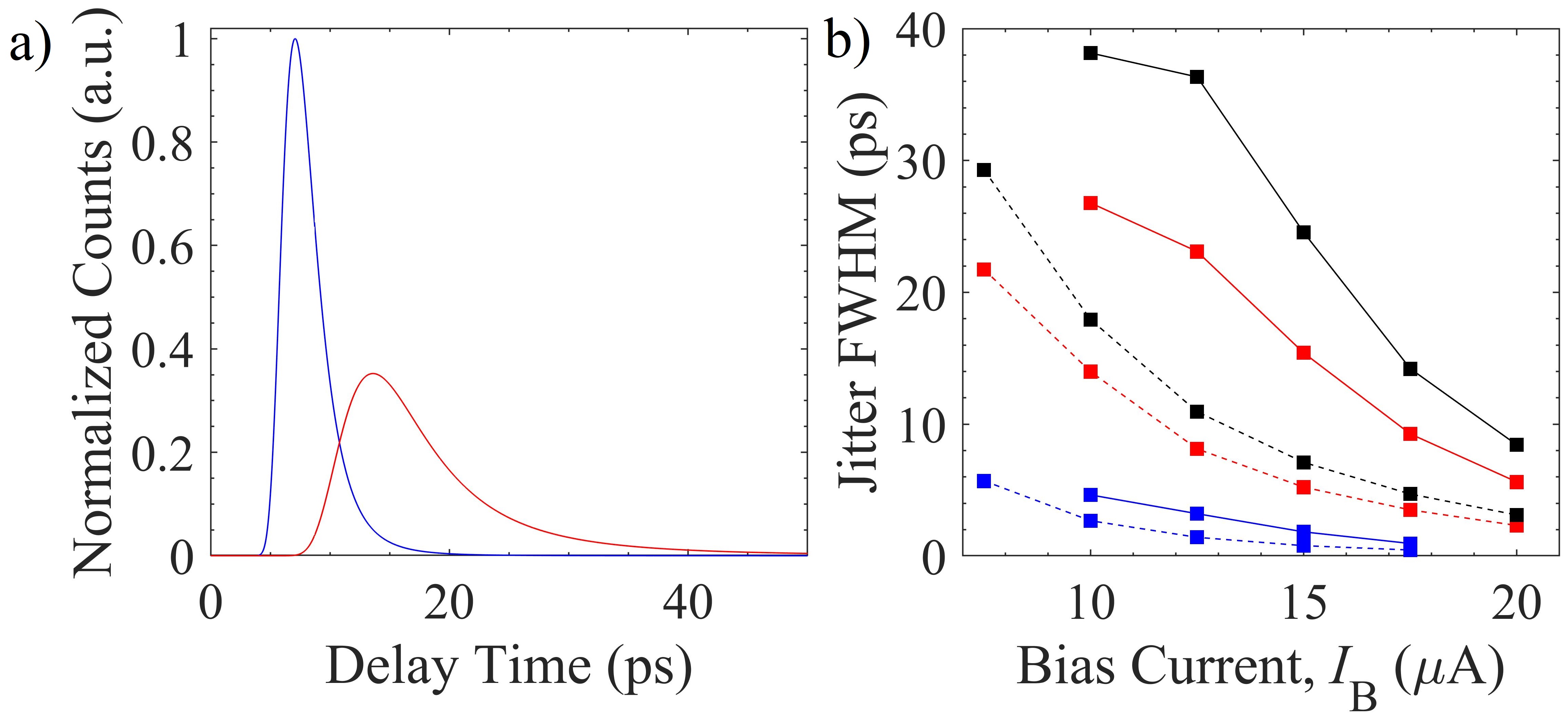}
\caption{(a) IRF for a bias current of 17.5 \(\mu\)A and \(\tau_{ee}\left(T_c\right) \) of 5 ps for 1550 nm (red) and 775 nm (blue) photon energies.  (b) Jitter FWHM vs. bias current for 1550 nm (solid) and 775 nm (dashed) photon energies.  The results are shown for \(\tau_{ee}\left(T_c\right) \) values of 0 ps (blue), 5 ps (red), and 10 ps (black).  The case of \(\tau_{ee}\left(T_c\right) \) reduced to 0 ps corresponds to the standard TDGL formulation, which follows from (\ref{G TDGL}) in the limit $\tau_{ee}\rightarrow0$. 
}
\label{F Hist FWHM}
\end{figure}

\quad The resulting latency fit can be used to generate the detector IRF according to (\ref{H t y 1}) as shown in Fig. \ref{F Hist FWHM} (a).  The FWHM of the jitter profile is extracted and shown in Fig. \ref{F Hist FWHM} (b) for 1550 nm and 775 nm photons for different values of the parameter \(\tau_{ee}\left(T_c\right) \).  At temperatures below \(T_c\), as experienced in the hotbelt model, \(\varrho\) is dominated by the electron-electron inelastic scattering time.  This parameter has the dominant influence on the latency compared to other parameters such as \(\tau_{esc}\) or \(\tau_{ep}\left(T_c\right) \).  Increasing \(\tau_{ee}\left(T_c\right) \) leads to slower suppression of the order parameter and significantly longer detection latency.  In the presence of energy fluctuations, this leads to an increase in the timing jitter.  As can be seen in Fig. \ref{F Hist FWHM} (b), the generalized TDGL equations show significantly different quantitative behavior when compared to the standard TDGL equations represented by \(\tau_{ee}\left(T_c\right) = 0\).  This correction is necessary to simulate intrinsic jitter on the same scale as observed experimentally \cite{korzh_demonstrating_2018}.  The fitting parameter \(\chi\) is selected to fit the $PCR$ curve of the experimental data for 1550 nm photons \cite{korzh_demonstrating_2018}, and kept constant for each value of \(\tau_{ee}\left(T_c\right) \).  The values of \(\chi \) for \(\tau_{ee}\left(T_c\right) \) of 0, 5, and 10 ps are 0.16, 0.39, and 0.52 respectively (compare to \(\chi=0.43 \) for the simple HB model of Section \ref{sec:SimplfiedHBModel}).

\begin{figure}
\includegraphics[width=\linewidth]{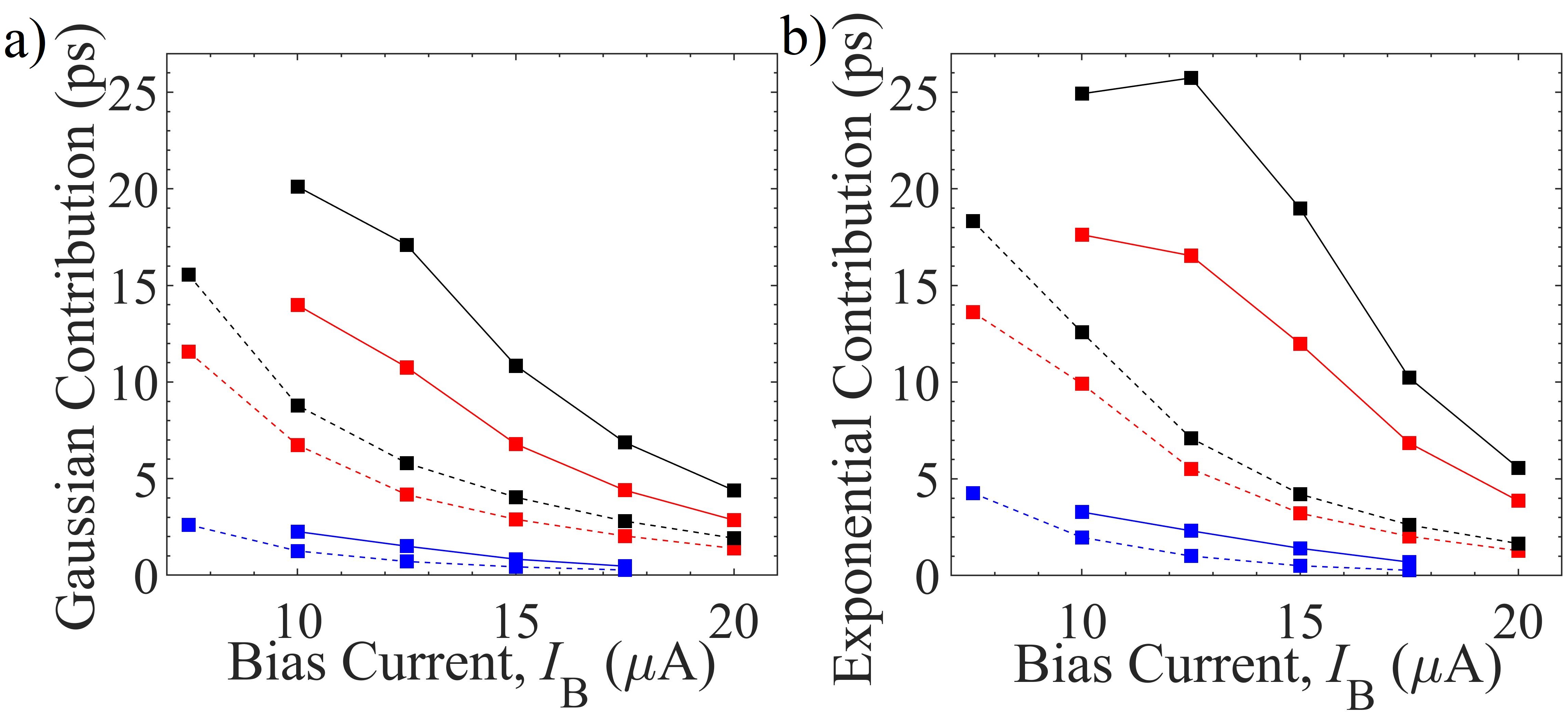}
\caption{Gaussian (a) and exponential (b) contributions to the timing jitter as a function of bias current for 1550 nm (solid) and 775 nm (dashed) photon energies when fitting the IRF with an exponential modified Gaussian function.  The results are shown for \(\tau_{ee}\left(T_c\right) \) values of 0 ps (blue), 5 ps (red), and 10 ps (black). 
}
\label{F Gaussian Exp}
\end{figure}

\quad The instrument response function is often fit with an exponentially modified Gaussian (EMG) profile to account for a notable tail observed experimentally \cite{najafi_timing_2015,sidorova_physical_2017,korzh_demonstrating_2018}.  The histograms generated using the generalized TDGL hotbelt formulation are not strictly defined by an EMG, but share the same characteristics of a mostly Gaussian profile with a tail at longer latency times.  As a comparison, we fit our simulation IRFs with an EMG distribution and extract the Gaussian FWHM (2.355\(\sigma\)) and exponential (\(1/\lambda\)) contribution as done in \cite{korzh_demonstrating_2018}.  Fig. \ref{F Gaussian Exp} shows the relative contribution of the Gaussian and exponential components of the fit.  The contributions are similar, with the exponential portion contributing slightly more to the timing jitter.  This is qualitatively consistent with the experimental findings of \cite{korzh_demonstrating_2018}.

\quad Experimental measurements of the intrinsic IRF may provide a means of distinguishing between the contributions of Fano fluctuations and spatial non-uniformities to the timing jitter.  Within the generalized 1D TDGL model, we use (\ref{TDGL model differential latency}) to calculate \({\left|\frac{\partial\tau_{lat}\left(y, I_B,T_b,\bar{E}\right)}{\partial \bar{E}} \right|}_{\bar{E}=\chi E_\lambda }\), which determines the FWHM of the Gaussian component of the timing jitter.  We obtain
\begin{equation}\label{tau lat derivative}
\begin{aligned}
\left|\frac{\partial\tau_{lat}\left(y, I_B,T_b,\bar{E}\right)}{\partial \bar{E}} \right|_{\bar{E}=\chi E_\lambda} = {}&\\
\big(\tau_{lat}\left(I_B,T_b,E\right) - \tau_{lat}&\left(I_B,T_b,\infty\right)\big)\times \\
\Bigg(\frac{\alpha\left(I_B,T_b\right)}{E - E_{det}\left(I_B, T_b\right)} &{}+ \frac{s \left(I_B,T_b\right)}{E}\Bigg).
\end{aligned}
\end{equation}
Taking \(I_B = I_{sw}\) and neglecting \(E_{det}\left(I_{sw},T_b\right) \ll E\) for dominant energy depositions, we obtain the asymptotic behavior at high photon energies near the switching current
\begin{equation}\label{FWHM asymptotic}
\begin{aligned}
\Upsilon \sim \sigma\left(E\right)E^{-\left(\alpha + s + a\right)}\sim
\left\lbrace 
\begin{array}{ll}
                  E^{-\left(\alpha+s+1/2\right)} \text{, Fano}\\
                  E^{-\left(\alpha+s\right)} \text{, non - uniformity}\end{array}
\right.
\end{aligned}     
\end{equation}
where \(\alpha =\alpha \left(I_{sw},T_b\right) \) and  \(s=s\left(I_{sw},T_b\right) \).  If instrumental sources of jitter are significantly lower than the intrinsic jitter, this difference in power law due to the differing energy scaling of Fano fluctuations and non-uniformity fluctuations may be apparent in the measured IRFs.  This method requires that an accurate model exists to describe the scaling of the latency with energy. 
\subsection{\label{sec:level2}Photon pair latency difference}
\begin{figure}
\includegraphics[width=0.8\linewidth]{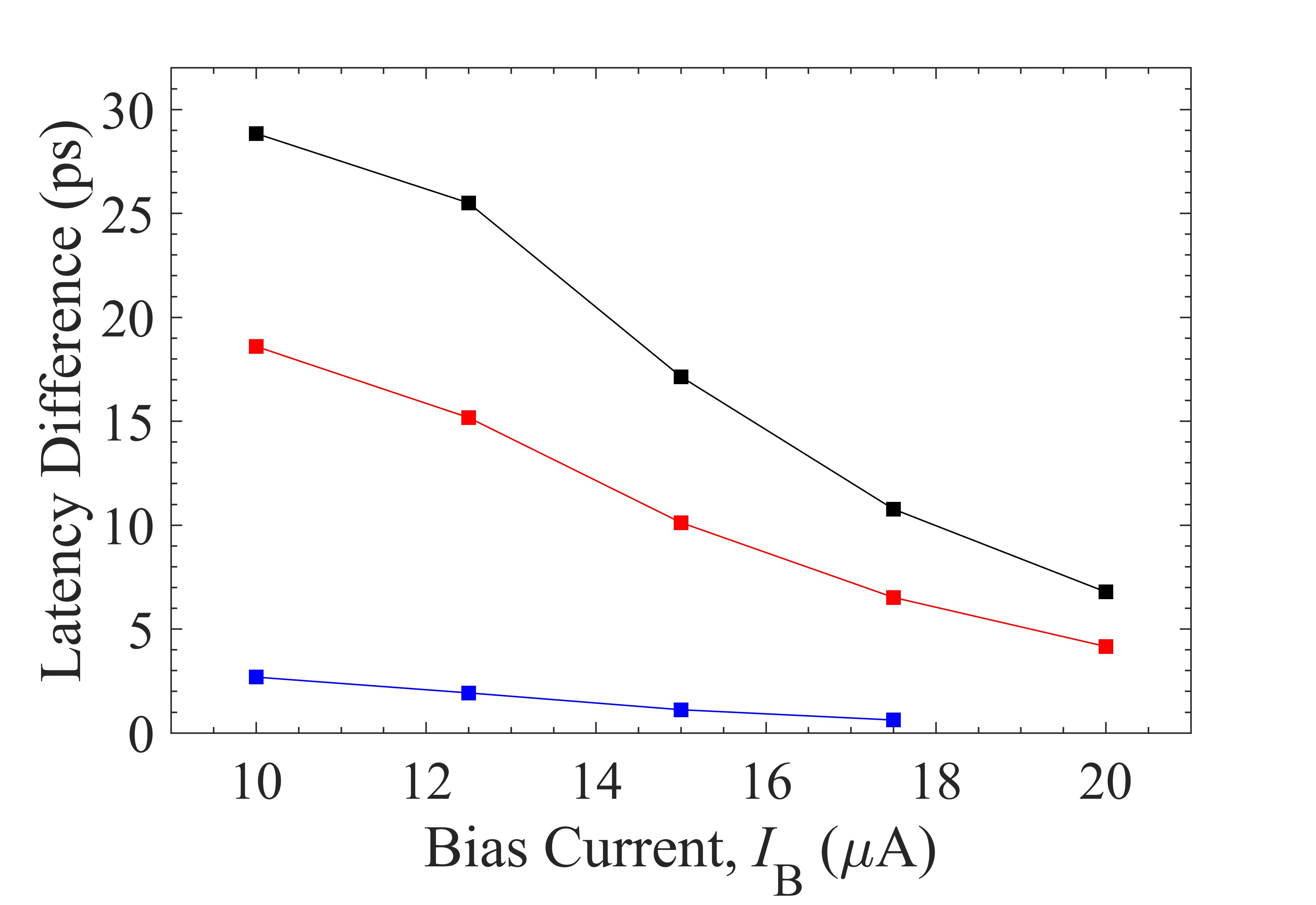}
\caption{Latency difference between photons of 1550 and 775 nm wavelength.  The results are shown for \(\tau_{ee}\left(T_c\right) \) values of 0 ps (blue), 5 ps (red), and 10 ps (black).  Increasing \(\tau_{ee}\left(T_c\right) \) leads to larger differential latency. 
}
\label{F Latency Difference}
\end{figure}

\quad Recent experiments \cite{korzh_demonstrating_2018} show that the latency difference between pairs of photons with 1550 nm and 775 nm wavelengths ranges from 5 to 25 ps depending on the bias current.  The latency difference is extracted from simulated IRFs by calculating the time difference between the maxima of the histograms of photon energies of 1550 and 775 nm.  Fig. \ref{F Latency Difference} shows the extracted latency difference for various bias currents and values of the parameter \(\tau_{ee}\left(T_c\right) \).  The latency difference shows a decreasing trend as the bias current increases, which is consistent with experiments.  Only by moving to the generalized TDGL formulation with non-zero \(\tau_{ee}\left(T_c\right) \) is it possible to find latency differences on the order of 5 to 25 ps as measured experimentally \cite{korzh_demonstrating_2018}.

\section{\label{sec:Discussion}Discussion}
\quad In an SNSPD, the four main stages governing the detector response are a) the initial equilibration in the system of interacting electronic excitations and phonons; b) a non-dissipative stage over which the system evolves with the order parameter being gradually suppressed until the superconducting state becomes incapable of supporting the current through the wire as a supercurrent.  In this stage, instability results in the generation of phase slip lines or vortices, depending on the symmetry of current flow through the region of suppressed order parameter (i.e. HB or HS regime); c) a dissipative stage resulting in the nucleation and expansion of the normal domain; and d) current diversion into the readout and relaxation of the normal domain. Stages a) to c) contribute to the intrinsic timing jitter of SNSPDs. 

\begin{figure}
\includegraphics[width=0.75\linewidth]{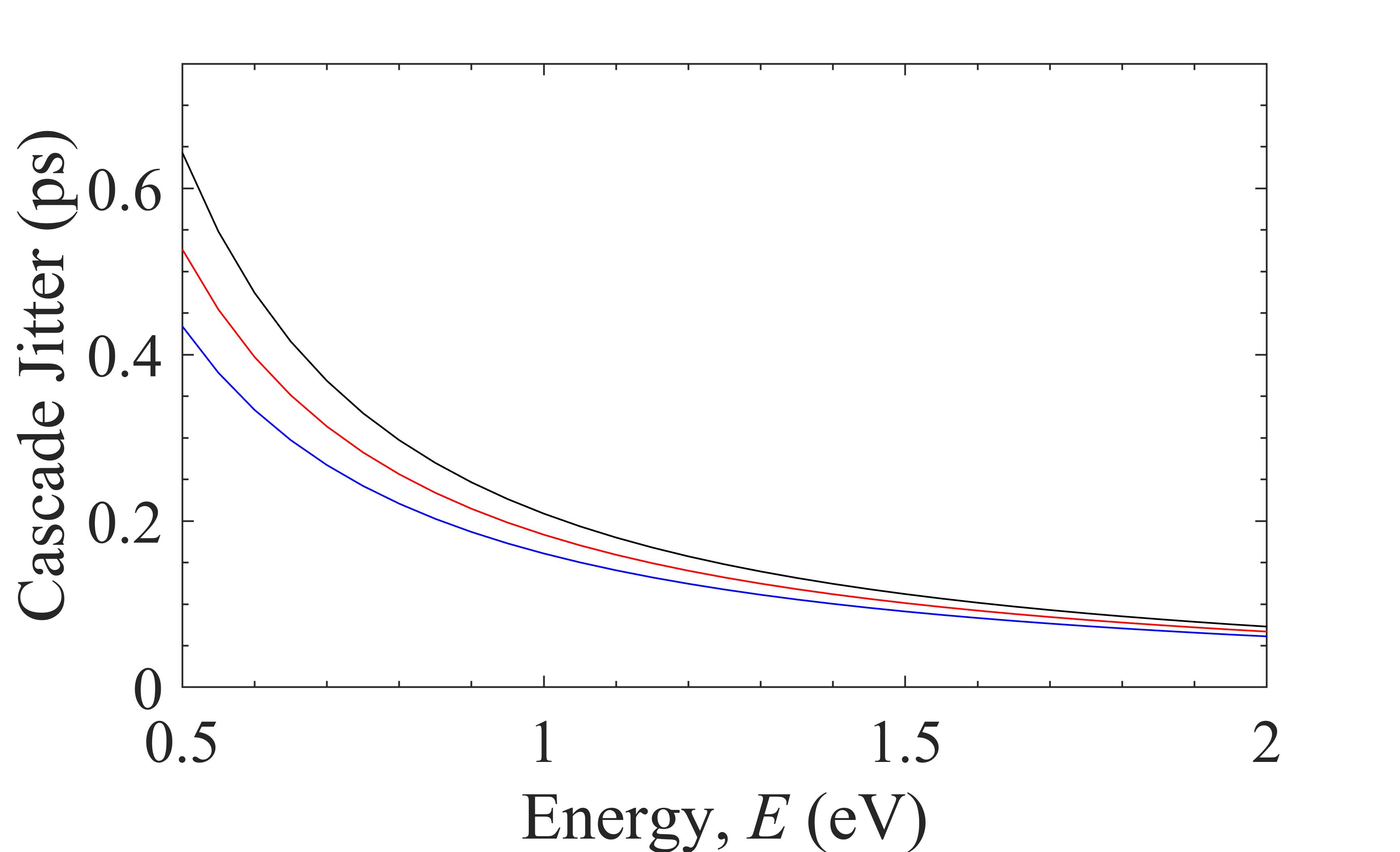}
\caption{Down-conversion cascade timing jitter estimate for NbN. Calculations use the initial hotspot volume \(V_{init} = 2\xi_c \times 2\xi_c d\) \cite{vodolazov_single-photon_2017} and \(\beta = 0.1\) (black), 1 (red), and 2 (blue).  Larger \(\beta\) reduces the cascade jitter \(\Upsilon_{th}(E)\). As the photon energy increases, the cascade jitter decreases.
}
\label{F Cascade Jitter}
\end{figure}

\quad During stage a) both Fano fluctuations and spatial non-uniformities cast their imprints onto the duration of stage b). To derive a simple expression for \(\delta\tau_{th}\) at the end of stage a), we write \(1/\tau_{th}(T_e) =  1/\tau_{ep}(T_e) + 1/\tau_{ee}(T_e)\) incorporating both electron-phonon and electron-electron scattering. The mean electron (and phonon) temperature after thermalization is found from the energy conservation law (\ref{conservation law}).  Due to Fano fluctuations the amount of energy deposited into the film fluctuates, which causes fluctuations in \(T_e\).  Differentiating (\ref{conservation law}) we have 

\begin{equation}\label{differentiated conservation law}
\frac{\delta E}{\mathcal{E}_0 V_{init}} = \frac{\delta T_e}{T_c}\frac{\pi^2}{3}\frac{T_e}{T_c}\left[\frac{1}{2} + \frac{4 \pi^2}{5 \gamma}\left(\frac{T_e}{T_c}\right)^2\right].
\end{equation}
After straightforward calculations, we obtain the standard deviation \(\sqrt[]{\overline{\left(\delta T_e\right)^2}}\) of the form 

\begin{equation}\label{temperature fluctuations}
\frac{\sqrt[]{\overline{\left(\delta T_e\right)^2}}}{T_c} = \frac{\sigma}{\mathcal{E}_0 V_{init}} \left\lbrace \frac{\pi^2}{3}\frac{T_e}{T_c}\left[\frac{1}{2} + \frac{4 \pi^2}{5 \gamma}\left(\frac{T_e}{T_c}\right)^2\right]\right\rbrace^{-1}
\end{equation}
and correspondingly

\begin{equation}\label{temperature fluctuations}
\begin{aligned}
&\sqrt[]{\overline{\left(\delta \tau_{th}\right)^2}} = \left|\frac{d\tau_{th}}{dT_e}\right|\sqrt[]{\overline{\left(\delta T_e\right)^2}} = \\
&3\tau_{ep}(T_c)\frac{\sigma}{\mathcal{E}_0 V_{init}}\frac{1 + \beta/3x^2}{\left(1 + \beta/x^2\right)^2}\left\lbrace \frac{\pi^2}{3}x^5\left[\frac{1}{2} + \frac{4 \pi^2}{5 \gamma}x^2\right]\right\rbrace^{-1}
\end{aligned}
\end{equation}
where \(\beta = \tau_{ep}(T_c)/\tau_{ee}(T_c)\) and \(x = T_e/T_c\). The contribution to the timing jitter becomes \(\Upsilon_{th} \approx 2.355~\sqrt[]{\overline{\left(\delta \tau_{th}\right)^2}}\). and is shown in Fig. \ref{F Cascade Jitter}. As seen from this estimate, the cascade contribution \(\Upsilon_{th}\) for typical NbN material parameters is in the sub-picosecond regime (\(\leq 0.3\)~ps for 1550~nm photons).  The only likely situation when the cascade duration may start contributing to the total latency \(\tau_{th} + \tau_{lat}\), and the cascade jitter to the total jitter, is when biasing the SNSPD at large currents close to the switching current.  In this case \(\tau_{lat}\) is expected to decrease with current, while the thermalization process at temperatures exceeding the critical temperature of a superconductor is practically independent of the current value. The main contribution to Fano-variance comes from losing phonons of the first few generations, where most of the deposited energy is in the form of phonons and their numbers are small \cite{kozorezov_fano_2017}. In contrast, the total amount of energy loss is accumulated over the whole latency time of the detector. The Fano fluctuations reflect the stochastic nature of phonon loss from the film.  The variance remains almost unchanged as the cascade enters the later stages of thermalization when approaching \(\tau_{th}\) (and beyond).  This is due to both the multiplication of phonon numbers in subsequent generations, so that the relative fluctuations for larger numbers become progressively smaller, and the gradual energy flow from phonons to electronic excitations while non-equilibrium state cools down.  For exactly the same reason, the cascade jitter is expected to be the same for both the HS and HB detection scenarios.

\quad To minimize the intrinsic timing jitter, the variances \({\sigma_{Fano}}^2\) and \({\sigma_{n-u}}^2\) must be reduced as much as possible.  Perfecting the technologies of superconducting thin film growth and nanowire fabrication are the principal ways to minimize the effects of spatial non-uniformities.  Their contribution to energy fluctuations relative to Fano fluctuations is expected to increase as the photon energy extends into the ultraviolet region.  Reducing the variance of Fano fluctuations generally requires reducing the escape of high-frequency phonons, including Debye phonons and the next generation of down-converted phonons. One way of implementing this is by increasing the thickness of the wire, making it as large as possible with respect to characteristic phonon mean free path \(d > l_{p-c} \).  Another approach is to control the acoustic properties of the escape interfaces between the nanowire and the substrate and the nanowire and the dielectric passivation layer. The latter may not work for acoustically soft metal films on rigid substrates, but may be of interest for acoustically fast NbN films, where higher frequency phonons may not escape if they face a gap in the phonon spectrum across the interface. It is important to emphasize that when the escape of high frequency phonons of the first generations of the down-conversion cascade has been greatly reduced or eliminated, lower frequency phonons, \(\hbar\omega<2\Delta\), may stay de-coupled from the condensate for longer than the duration of the latency and/or equilibration. Fluctuations in their number occur due to the energy partition between electrons and phonons and are described by the Fano variance \(\sigma^2 = F_{\text{eff}} \varepsilon E_{\lambda} \sim F_{\text{eff}} 1.75 \Delta E_{\lambda}\). Thus, in the ideal case the Fano-fluctuations effect can be reduced down to that described by the standard deviation \(\sigma_{Fano} \sim \sqrt[]{F_{\text{eff}} 1.75\Delta E_{\lambda}}\). This is nearly a factor of four smaller than the \(\sigma_{Fano}\) for 1550~nm used for simulations in Figs. \ref{F Histograms}-\ref{F HB Latency} and \ref{F Hist FWHM}-\ref{F Latency Difference}.  Such a dramatic decrease in the strength of fluctuations will move the intrinsic timing jitter (neglecting spatial inhomogeneities) into the sub-picosecond range.  Even if such ideal performance cannot be easily achieved, the potential for improvement is significant, and further work is needed in acoustic matching and the control of interface properties and phonon fluxes including internal phonon bottle-necking in acoustically soft metal films on rigid substrates and with rigid passivation layers \cite{sidorova_electron-phonon_2016}.    

\quad Stage b) comprises the longest part of the detector latency, during which the cooling of the affected area due to phonon escape and thermal conduction is accompanied by the suppression of the order parameter, resulting in an instability of the superconducting state and transition into the resistive state. The control parameters describe the energy exchange between quasiparticles and phonons, \(\tau_{ep}\), the exchange and loss of phonons to the thermal bath \(\tau_{esc}\) (as important in energy balance over the duration of stage b) as controlling the fluctuations in stage a)), the thermal conductivity in the electronic system \(k_s\), and the energy relaxation time, \(\tau_{sc}\left(T_e\right)\).  

\quad As evidenced by our analysis and pointed out in Section \ref{sec:TDGLModel}, the strongest potential impact on reducing the latency and jitter is to make the suppression of superconductivity as fast as possible.  We modeled the presence of an extra channel of inelastic scattering by adding the electron-electron scattering in parallel with the electron-phonon interaction.  However, in a disordered 2D metal film, the energy transfer in the diffusion channel is small, making the energy relaxation in electron-electron collisions slower than the phase relaxation time \cite{a._l._efros_electron-electron_1985}.  The thermalization time was reported as 7~ps in NbN \cite{ilin_ultimate_1998}, which is consistent with the prediction of Altshuler-Aronov formula \cite{korneeva_comparison_2017}.  The relatively weak effect of electron-electron scattering was incorporated in the form of an electron-electron collision integral in the kinetic theory of photon detection \cite{vodolazov_single-photon_2017}.  However, it is known that in disordered mesoscopic metal wires \cite{pierre_dephasing_2003}, dilute magnetic impurities less than 1 ppm can cause both anomalous decoherence and fast energy exchange.  Whether this is the case for the material used in recent experiments \cite{korzh_demonstrating_2018}, making it closer to optimum in terms of intrinsic jitter performance, is not immediately clear.  Nonetheless, the prospect of improving detector latency and timing jitter by introducing a controlled amount of spin-flip scattering, affecting the energy relaxation time, deserves further attention.

\quad In the 1D generalized TDGL model described in Section \ref{sec:TDGLModel}, the effect of the transverse coordinate dependence on detector latency is neglected.  This simplification is partially supported by recent measurements \cite{korzh_demonstrating_2018}.  When the IRF was measured for nanowire widths ranging from 60~nm to 120~nm, the timing jitter ranged from approximately 25~ps to 5~ps independent of the width, once the bias current was scaled according to the detection energy and width of the $PCR$ curves.  This suggests that the timing differences due to vortex crossing and the varying geometry of phase-slip lines as a function of transverse absorption coordinate may be much smaller than those due to energy fluctuations as described above.  If timing jitter due to transverse coordinate dependence were to dominate, the intrinsic jitter would be expected to show a more significant dependence on nanowire width, because the vortex transit times would change.  Furthermore, the estimates of the timing jitter due to vortex crossing based on a model of vortex entry from the nanowire edge suggest that this effect is small compared to the measured intrinsic jitter \cite{wu_vortex-crossing-induced_2017}.  Note that this work neglected the effect of vortex-antivortex unbinding due to detections near the center of the nanowire, which will alter the expected IRF. Consideration of vortex-antivortex unbinding inside the hotspot must be performed using the 2D generalized TDGL model, which is left for further analysis. However, we notice that for all data available at this time, as evidenced by the period of voltage oscillations (see our Fig. \ref{F Vtrace} and Fig. 8 of \cite{vodolazov_single-photon_2017}), that vortex crossing times appear to be too short in comparison with the observed latencies of the detector to be dominant. Therefore, the contribution of these effects to timing jitter must be of minor importance. 

\quad The measurement of detection latency differences on the order of 25~ps to 5~ps between 1550~nm and 775~nm photons further supports the use of a 1D model, because the absolute latency must be of the same magnitude.  The characteristic diffusion time to establish a hotbelt based on the parameters of NbN used in Section \ref{sec:TDGLModel} is \(\tau_{D,W} \simeq w^2/16D - w^2/4D  \simeq 8 \text{ ps} - 32 \text{ ps}\) for \(W=80 \text{ nm}\), which is comparable to the measured latency.  The generalized TDGL model captures this behavior through the introduction of the parameter \(\varrho\) which modifies the characteristic time of the variation of the order parameter under the standard TDGL model, \(\tau_{\left|\Delta\right|}\).  Under this modification, the condition for quasi-1D detection, \(\varrho \tau_{\left|\Delta\right|} \simeq \tau_{D,W}\), can be satisfied  even when calculation according to the standard TDGL formation suggests that a 1D model is inappropriate because \(\tau_{\left|\Delta\right|} \ll \tau_{D,W}\).  As the bias current approaches \(I_{sw}\), and the latency falls below the characteristic diffusion time and the 1D approximation becomes less appropriate.  In this regime, the effects of transverse coordinate dependence may become important.  More advanced modeling in the future, using the 2D generalized TDGL formulation, could provide information about the impact of the transverse coordinate of absorption relative to energy fluctuations.  The inclusion of the full 2D dynamics is also expected to change the behavior of the longer latency tail observed experimentally and through modeling.  Detailing the nature of this non-Gaussian behavior may be another way to experimentally probe the details of the detection mechanism.

\quad Within our 1D model, stage c) is less significant.  Due to the uniform nature of the current flow within the cross-section, the normal domain expands at the same rate for excitations of different energy at the same bias current.  In the 2D case, the initial stage of the growth of the normal domain may contribute to timing jitter because the expansion depends on the initial coordinate of detection.  In this case, the resulting timing jitter would be correlated with the transverse coordinate dependence caused by vortex nucleation and motion.  While not important within the model of intrinsic jitter, the rate of normal domain growth plays a role in an experimental system where electrical noise contributes to timing jitter.  The slew rate of the rising edge of the electrical signal from a detection event determines the timing jitter associated with the electrical noise in the system.  Maximizing the rate at which the normal domain grows and diverts current (stages c and d) can reduce the electrical noise contribution to the total timing jitter.

\quad Comparing the results of Sec. \ref{sec:TDGLModel} to recent experiments \cite{korzh_demonstrating_2018}, we find a good qualitative match for a value of \(\tau_{ee}\left(T_e\right) \sim 5-10 \text{ ps}\) which is consistent with the value of \(\tau_{ee}\left(T_e\right) \) predicted by experiment \cite{ilin_ultimate_1998} and theory \cite{korneeva_comparison_2017}.  The detection latency difference between pairs of 1550 and 775~nm photons shows the same monotonically decreasing behavior with increasing bias current and similarly shows an inflection within the range of bias currents that do not saturate the internal detection efficiency.  The qualitative behavior of the IRF FWHM also matches the experiment for the same values of the inelastic scattering time.  In both experiment \cite{korzh_demonstrating_2018} and the model, the IRF FWHM for 1550~nm photons decreases monotonically with an inflection point occurring at a bias current just below the current where internal efficiency saturates.  For 775~nm photons, no inflection point is observed.  The simple model predicts a somewhat higher exponential contribution to the total jitter compared to the Gaussian contribution than found experimentally, but the qualitative behavior of the two components each match experiment.  Despite its 1D approximation, the generalized TDGL model captures all of the qualitative behavior observed for the 80 nm sample measured in \cite{korzh_demonstrating_2018}.  A better quantitative match to experiments could be found by using \(\tau_{ee}\left(T_c\right)\) as a fitting parameter, but that is outside the scope of this manuscript.

\section{\label{sec:Conclusion}Conclusion}
\quad We have demonstrated how the introduction of detector latency in the presence of Fano fluctuations and spatial inhomogeneities accurately reproduces the qualitative features of the intrinsic timing jitter recently measured in SNSPDs.  In the characteristic latency vs. energy curve, the presence of a singularity at the detection energy combined with monotonic scaling and positive curvature leads to a non-Gaussian IRF with an extended tail at longer delay times, which was observed in the experiment.  Within the framework of the generalized TDGL model, the inelastic scattering time plays a dominant role in determining the detector latency and timing jitter.  The addition of this contribution to the standard TDGL model is necessary to reproduce the detector latency observed in experiment.  By engineering materials with smaller \(\tau_{sc}\left(T_e\right)\), it may be possible to reduce this component of the intrinsic jitter in the future.  The structural features of detector IRFs such as FWHM, asymmetry characteristics, and latency differences between pairs of photons with different energies offer a new means of studying the detection mechanism in SNSPDs in more detail going forward.

\section*{\label{sec:Acknowledgements}Acknowledgements}
\quad Part of this research was performed at the Jet Propulsion Laboratory, California Institute of Technology, under contract with the National Aeronautics and Space Administration.  Support for this work was provided in part by the DARPA Defense Sciences Office, through the DETECT program.  J. P. A. was supported by a NASA Space Technology Research Fellowship.  The authors would like to thank S. Frasca and E. Ramirez for sharing unpublished experimental results and S. Young, M. Sarovar, F. Leonard, J. Bienfang, and S. W. Nam for helpful discussions. 

\nocite{apsrev41Control}
\bibliographystyle{apsrev4-1_ModStyle}
\bibliography{SNSPD_Theory_Cleaned.bib}

\end{document}